\documentclass[longauth]{aa}

\usepackage{graphicx}
\usepackage[varg]{txfonts}
\usepackage{longtable}
\usepackage[T1]{fontenc}
\usepackage{ae,aecompl}
\usepackage{natbib}
\bibpunct{(}{)}{;}{a}{}{,} 
\usepackage{amsmath}	
\usepackage{amssymb}	
\usepackage{tikz,graphics,color,float,epsf} 
\usepackage{lscape} 
\usepackage{rotating}
\usepackage{xcolor}
\usepackage{calc}
\usepackage{hyperref}
\usepackage{longtable}
\usepackage[english]{babel}
\usepackage{lipsum}

\newcommand{\lsun}{\mbox{L}_\odot}
\newcommand{\rsun}{\mbox{R}_\odot}
\newcommand{\msun}{\mbox{M}_\odot}
\newcommand{\pab}{\mbox{Pa}\beta}
\newcommand{\brg}{\mbox{Br}\gamma}
\newcommand{\lacc}{L_{\rm acc}}
\newcommand{\macc}{\dot{M}_{\rm acc}}
\newcommand{\lstar}{L_\star}
\newcommand{\mstar}{M_\star}
\newcommand{\rstar}{R_\star}
\newcommand{\teff}{T_{\rm eff}}
\newcommand{\lbol}{L_{\rm bol}}

\begin{document}

   \title{The Enigma of Gaia18cjb: a Rare Hybrid of FUor and EXor?
   \\
   \thanks{Based on observations collected at the Large Binocular Telescope under LBT programme IT-2019B-008.} 
   \thanks{Based on observations collected at the European Southern Observatory under ESO/NTT programmes 105.203T.001 and 105.203T.003.}
   \thanks{Based on observations collected at the Gran Telescopio Canarias under GTC program GTC29-22B.}
}

\titlerunning{Gaia18cjb: a Hybrid of FUor and EXor?}


\author{Eleonora Fiorellino\inst{1,2,3} \and 
P\'eter \'Abrah\'am\inst{2,3,4,5}  \and \'Agnes K\'osp\'al\inst{2,3,4,6} \and M\'aria Kun\inst{2,3} \and Juan M. Alcalà\inst{1}  \and Alessio Caratti o Garatti\inst{1} \and Fernando Cruz-S\'aenz de Miera\inst{2,3,7} \and  David García-Álvarez\inst{8,9} \and  Teresa Giannini\inst{10} \and  Sunkyung Park\inst{2,3} \and Micha{\l} Siwak\inst{2,3} \and M\'at\'e Szil\'agyi\inst{2,3,4}  
\and Elvira Covino\inst{1}  \and Gabor Marton\inst{2,3} \and Zs\'ofia Nagy\inst{2,3} \and Brunella Nisini\inst{10} \and Zs\'ofia Marianna Szab\'o\inst{2,3,11,12}
\and  Zs\'ofia Bora\inst{2,3,4}  \and Borb\'ala Cseh\inst{2,3,13} \and Csilla Kalup\inst{2,3,4} \and M\'at\'e Krezinger\inst{2,3,4}
\and Levente Kriskovics\inst{2,3} \and Waldemar Og{\l}oza\inst{14} \and Andr\'as P\'al\inst{2,3} \and \'Ad\'am S\'odor\inst{2,3}
\and Eda Sonbas\inst{15,16} \and R\'obert Szak\'ats\inst{2,3} \and Kriszti\'an Vida\inst{2,3,4} \and J\'ozsef Vink\'o\inst{2,3}  
\and Lukasz Wyrzykowski\inst{17} \and Pawel Zielinski\inst{18} }
\institute{INAF-Osservatorio Astronomico di Capodimonte, via Moiariello 16, 80131 Napoli, Italy\\
\email{eleonora.fiorellino@inaf.it}
\and Konkoly Observatory, HUN-REN Research Centre for Astronomy and Earth Sciences, Konkoly-Thege Mikl\'os \'ut 15-17, 1121 Budapest, Hungary
\and CSFK, MTA Centre of Excellence, Konkoly-Thege Mikl\'os \'ut 15-17, 1121 Budapest, Hungary
\and ELTE E\"otv\"os Lor\'and University, Institute of Physics, P\'azm\'any P\'eter s\'et\'any 1/A, 1117 Budapest, Hungary
\and Institute for Astronomy (IfA), University of Vienna, T\"urkenschanzstrasse 17, A-1180 Vienna, Austria
\and Max Planck Institute for Astronomy, K\"onigstuhl 17, 69117 Heidelberg, Germany
\and 
Institut de Recherche en Astrophysique et Plan\'etologie, Universit\'e de Toulouse, UT3-PS, OMP, CNRS, 9 av.\ du Colonel Roche, 31028 Toulouse Cedex 4, France
\and 
Grantecan S.A., Centro de Astrofísica de La Palma, Cuesta de San José, E-38712 Breña Baja, La Palma, Spain
\and Instituto de
Astrofísica de Canarias, Avenida Vía Láctea, E-38205 La Laguna, Tenerife, Spain
\and INAF-Osservatorio Astronomico di Roma, via di Frascati 33, 00078, Monte Porzio Catone, Italy
\and Max-Planck-Institut f\"ur Radioastronomie, Auf dem H\"ugel 69, 53121 Bonn, Germany
\and Scottish Universities Physics Alliance (SUPA), School of Physics and Astronomy, University of St Andrews, North Haugh, St Andrews, KY16 9SS, UK
\and MTA-ELTE Lend{\"u}let "Momentum" Milky Way Research Group, Hungary
\and Mount Suhora Astronomical Observatory, Cracow Pedagogical University, ul. Podchorazych 2, 30-084 Krak{\'o}w, Poland
\and Adiyaman University, Department of Physics, 02040 Adiyaman, Turkey
\and
Astrophysics Application and Research Center, Adiyaman University, Adiyaman 02040, Turkey
\and
Astronomical Observatory, University of Warsaw, Al.~Ujazdowskie~4, 00-478~Warszawa, Poland
\and 
Institute of Astronomy, Faculty of Physics, Astronomy and Informatics, Nicolaus Copernicus University in Toru{\'n}, Grudzi\k{a}dzka 5, 87-100 Toru{\'n}, Poland
}
   \date{Received ; accepted }

\abstract
{Gaia18cjb is one of the Gaia-alerted eruptive young star candidates which has been experiencing a slow and strong brightening during the last 13 years, similar to some FU Orionis-type objects.} 
{The aim of this work is to derive the young stellar nature of Gaia18cjb, determine its physical and accretion properties to classify its variability.} 
{We conducted monitoring observations using multi-filter optical and near-infrared photometry, as well as near-infrared spectroscopy. 
We present the analysis of pre-outburst and outburst optical and infrared light curves, color-magnitude diagrams in different bands, the detection of near-IR spectral lines, and estimates of both stellar and accretion parameters during the burst.}
{The optical light curve shows an unusually long (8 years) brightening event of 5\,mag in the last 13 years, before reaching a plateau indicating that the burst is still on-going, suggesting a FUor-like nature. 
The same outburst is less strong in the infrared light curves.
The near-infrared spectra, obtained during the outburst, exhibit emission lines typical of highly accreting low-intermediate mass young stars with typical EXor features. 
The spectral index of Gaia18cjb SED classifies it as a Class I in the pre-burst stage and a Flat Spectrum young stellar object (YSO) during the burst. 
}
{
Gaia18cjb is an eruptive YSO which shows FUor-like photometric features (in terms of brightening amplitude and length of the burst) and EXor-like spectroscopic features and accretion rate, as V350\,Cep and V1647\,Ori, classified as objects in between FUors and EXors. 
}
\maketitle
%

\section{Introduction} \label{sect:intro}

Low mass ($< 2M_\odot$) young stellar objects (YSOs) exhibit photometric variability in the optical and infrared bands on timescales spanning from minutes to centuries \citep{car01, meg12, cody2014, siwak2018}. Their photometric variability can be caused by changes in accretion rate, varying line-of-sight extinction, or rotating accretion hot or cold spots \citep[see][for a review]{fis22pp7}. 

Among variable sources, the so called eruptive young stars (EYSs) show the largest brightness increases. 
Historically, we classified EYSs on the amplitude of the brightness change. 
EX\,Lupi-type objects \citep[EXors,][]{Herbig89} present an {\it EXor burst} of about $1-2.5$\,mag within timescales ranging from weeks up to 1\,year in the framework of the magnetospheric accretion scenario. 
Their spectra are very similar to Class\,II pre-main sequence stars (PMSs) or Classical T\,Tauri stars (CTTSs). 
FU\,Orionis-type objects \citep[FUors,][]{Herbig77}, on the other hand, present a stronger brightening, of $ 2.5 - 6$\,mag, and it takes months to years to reach the peak. 
Spectroscopically, FUors show absorption-line profiles, formed because the disc atmosphere is heated from the mid-plane, and their spectral types depend on the observed wavelengths: F$-$G type in optical and K$-$M type in near-infrared \citep[NIR;][]{har96, audard2014, connelley-reipurth2018, fis22pp7}. 
Most importantly, the nature of the FUor outburst cannot be explained by magnetospheric accretion and it is likely to be related to other accretion mechanisms, as for example the boundary layer accretion, where the mass accretion rate ($\macc$) is so high that shrinks the stellar magnetic field down to the stellar radius. 
Whichever the mechanism and the order of magnitude, both EXor and FUor outbursts fuel the build-up of the protostellar mass, increasing dramatically the mass accretion rate from the disk to the central forming star.
Such strong bursts are thought to be a possible explanation to the protostellar luminosity spread \citep{ken90, eva09, fis22pp7}, consisting in the fact that YSOs observed luminosity is larger by about an order of magnitude than what was expected by the standard steady-state collapse mode by \citet{shu77}. 
\par Recent works show that some outbursting YSOs have peculiar spectroscopic properties and light-curves that do not fit any of the aforementioned two classical categories \citep[see][and references therein]{fis22pp7}. 
Detailed studies of individual objects are important to understand the physical mechanism, the evolution, and the properties of very different EYSs, and to understand how eruptive accretion impacts the protostellar mass building process and the interplay with the steady magnetospheric accretion process.

In the last decade, also thanks to the whole-sky monitoring of the Gaia space observatory and its Gaia Photometric Science Alerts Program\footnote{\url{http://gsaweb.ast.cam.ac.uk/alerts/home}} \citep{Hodgkin2021}, many EYSs have been discovered. These new EYSs have been classified as FUors, such as Gaia17bpi \citep{hillenbrand2018}, Gaia18dvy \citep{Szegedi-Elek2020}, Gaia21bty \citep{Siwak2023}, and Gaia21elv \citep{nagy2023}; EXors, like Gaia18dvz \citep{Hodapp2019}, Gaia19fct \citep{Park2022}, and Gaia20eae \citep{Cruz-SaenzdeMiera2022, Ghosh2022}; and in between, such as Gaia19ajj \citep{hillenbrand2019}, and Gaia19bey \citep{hodapp2020}.

Our study of Gaia18cjb enters in this context. An alert was issued for this source by Gaia Photometric Science Alerts Program on 2018 August 21. After the alert, we started monitoring this source with photometric and spectroscopic follow-up observations. 

Based on information from Gaia DR2, the Wide-field Infrared Survey Explorer (WISE), and Planck measurements,  \citet{Marton2019} evaluated that the probability of Gaia18cjb being a YSO is 83\%.
It is located at (R.A., Dec.) = (06h39m07.54s,   00d08m54.49s). 
The Gaia DR3 measured a parallax of $p = 0.83 \pm 0.69$\,mas, corresponding to a distance range between 660\,pc and 7140\,pc. 
\citet{bai21} proposed a prior-dominated distance estimate for sources with large uncertainties on their parallax, finding Gaia18cjb can be located at a distance of $d=1.03^{+0.20}_{-0.19}$\,kpc. However, this determination uses a prior for the absolute G-band which is dominated by Main-Sequence stars. The issue of Gaia18cjb's distance will be further discussed in Sect.\,\ref{App:distance}.

In this work we present the data we collected (Sect.\,\ref{sect:obs}), and our analysis of these data (Sect.\,\ref{sect:analysis}). We discuss the distance, the YSO nature of Gaia18cjb and its outburst on the basis of our results in Sect.\,\ref{sect:discussion}. Conclusions are highlighted in  Sect.\,\ref{sect:conclusions}.

\section{Observations and data reduction} \label{sect:obs}
To probe the eruptive nature of Gaia18cjb, we collected a plethora of photometric and spectroscopic new observations, and archival data with several telescopes. 
In this section we describe the dataset and its reduction.

\subsection{Photometry}

Technical details of the optical, infrared, and archival photometric data are listed in Table\,\ref{tab:photometry-text}.  
In the following subsections we describe in details the observations and the data reduction of the collected photometry.

\label{subsect:photometry}

\begin{table*}
\centering
\caption{\label{tab:photometry-text}Summary of photometric observations of Gaia18cjb.}
\begin{tabular}{llccccc}
\hline
\hline
Telescope & Dates & Camera & Bands & Pixel Scale & FoV \\ 
\hline		
\hline
RC80 & 2020 - 2023 & FLI PL230 CCD & $B$,$V$,$g'$,$r'$,$i'$ & 0$\farcs$55 & $18\farcm8 \times 18\farcm8$  \\
Suhora & 2021  & Apogee Aspen-47 & $B$,$V$,$g'$,$r'$,$i'$ & 1$\farcs$116 & $19\farcm0 \times 19\farcm0$  \\
ADYU60 & 2020 - 2021 & Andor iKon-M934 & $g'$,$r'$,$i'$ & 0$\farcs$673 & $11\farcm5 \times 11\farcm5$\\
NTT & 2021 & EFOSC & $B$,$V$,$R$ & 0$\farcs$24 & $4\farcm1 \times 4\farcm1$ \\ 
\hline
NTT & 2021  & SOFI & $J$,$H$,$K$ & 0$\farcs$273 & $4\farcm5 \times 4\farcm5$ \\ 
GTC & 2022 & EMIR & $Y$,$J$,$H$,$K$ & 0$\farcs$1945 & $6\farcm67 \times 6\farcm67$ & \\
\hline
ZTF & 2017 - 2023 & Cryogenic CCD & $g'$,$r'$ & 1$\farcs$0 & $47^\circ \times 47^\circ$\\
PanSTARRS & 2010 - 2014 & PS1 GPC1 &  $g'$,$r'$,$i'$,$z$,$Y$ & 0$\farcs$11 & $7^\circ \times 7^\circ$\\
2MASS & 2000 & 3 NICMOS3 & $J$,$H$,$K_s$ & 2$\farcs$0 & $8\farcm5 \times 8\farcm5$\\
UKIDSS & 2007 - 2010 & WFCAM & $J$,$H$,$K_s$ & 0$\farcs$4 & $0.21^\circ \times 0.21^\circ$\\
WISE & 2010 - 2023 & HAWAII 1RG & $W1$,$W2$ & 2.75$''$ & $47' \times 47'$\\

\hline
\hline
\end{tabular}
 \end{table*}

\begin{table*}
\centering
\caption{\label{tab:spectroscopy-text}Summary of spectroscopic observations of Gaia18cjb.}
\begin{tabular}{llcccc}
\hline
\hline
Telescope & Dates & Spectrograph & Spectral Range & Resolution \\
 & & & $\mu$m & \\
\hline		
\hline
LT & 2020 Nov & SPRAT & 0.4020 - 0.7994 & 350 \\
LBT & 2021 Jan & LUCI & 0.96 - 2.44 & 2000 \\
NTT & 2021 May & SOFI & 0.95 - 2.52 & 1000 \\
GTC & 2022 Oct & EMIR & 1.17 - 2.37 & 4000-5000 \\
\hline
\hline
\end{tabular}
 \end{table*}

\subsubsection{Optical Photometry}
We have been monitoring Gaia18cjb since October 2020 with an approximately monthly cadence using the 80\,cm Ritchey–Chrétien (RC80) telescope at the Piszkéstető mountain station of Konkoly Observatory (Hungary). The telescope is equipped with an FLI PL230 CCD camera 
and Johnson $BV$ and Sloan $g'r'i'$ filters. 
The aperture photometry for Gaia18cjb was obtained using 40 comparison stars located within a $12' \times 12'$ box around our science target. 
We used an aperture radius of 5 pixels ($2\farcs$75) and sky annulus between 20 and 40 pixels ($11''$ and $22''$). 
The instrumental magnitudes were converted to the standard system using the APASS9 magnitudes \citep{henden2016} of the comparison stars and fitting a linear color term.

Gaia18cjb was also observed on three nights between 2021 January and February at the Mount Suhora Observatory of the Cracow Pedagogical University (Poland) with the 60\,cm Carl–Zeiss telescope equipped with an Apogee Aspen-47 camera 
(Johnson $BV$ and Sloan $g'r'i'$ filters). 
We obtained aperture photometry using the same aperture radius and sky annulus (in arcseconds) as for the RC80 data, and same 40 comparison stars for the photometric calibration.

Further observations were collected on six nights between 2020 November and 2021 February at Adiyaman University using ADYU60 Application and Research Center using ADYU60, a PlaneWave 60\,cm f/6.5 corrected Dall–Kirkham Astrograph telescope. This telescope is equipped with an Andor iKon-M934 camera 
(Sloan $g'r'i'$ filters). Aperture photometry and calibration was done the same way as described above.

On 2021 May 9, we further observed Gaia18cjb with the 3.6\,m New Technology Telescope (NTT) located in La Silla Observatory (Chile). We used the ESO Faint Object Spectrograph and Camera \citep[EFOSC,][Bessell $BVR$ filters]{Arnaboldi2016}. 
Aperture photometry was obtained as above. For the photometric calibration, we used five stars from the set above that were included in the smaller FoV of the detector without fitting a color term. We calculated their Bessell $R$ magnitudes by interpolating in their spectral energy distribution using their APASS9 $BVg'r'i'$ fluxes.

In all the images that we acquired, the target looked point-like and there were no issue with extented emission.

\subsubsection{Infrared Photometry}  
\label{sec:sofi}
On 2021 May 9, we observed Gaia18cjb with the ESO New Technology Telescope (NTT) infrared spectrograph and imaging camera, i.e. SofI \citep{Arnaboldi2016}. 
The data reduction of the photometry was done as prescribed by SOFI manual\footnote{\url{https://www.eso.org/sci/facilities/lasilla/instruments/sofi/tools/reduction.html}}. 
We first constructed a sky frame by taking the median of the 5 dither positions after scaling them to the same mean level. The flatfield frame was constructed by subtracting a lamp off image from a lamp on frame and correcting for the residual shade pattern. The resulting flat frames show vertical stripes on the left hand quadrants.
To remove the stripes in the SW quadrant we applied an empirical correction, by taking the horizontal lines one-by-one from the affected upper-right quadrant, subtracted from each 512 pixel long horizontal line its median value; and added back to it the median of the remaining three quadrants of the whole image. 
After that, we divided the sky-subtracted and stripe-corrected frames by the flatfield.
From the 5 dither positions we constructed a mosaic per filter. 
Finally, we performed photometry on the mosaics. We used a 9 pixel radius aperture, with a sky ring between 7.8$''$ and 10.4$''$. 
For photometric calibration we used ~30 2MASS sources with quality flag ‘A’. 

We also obtained $JHK$-band photometry of Gaia18cjb on 2022 October 10th and 26th using the 10.4m Gran Telescopio Canarias (GTC) installed at the Spanish Observatorio del Roque de los Muchachos of the Instituto de Astrofísica de Canarias, on the island of La Palma, with the wide field Espectr\'ografo Multiobjeto Infra-Rojo (EMIR) Imager \citep{Garzon2022}. 
The images were reduced using dedicated python routines customised by the GTC staff for EMIR photometric data. The images were flat-fielded and the sky background was eliminated. The astrometric solution was also calculated. The final reduced image is the average of all the available images for each filter. We performed photometry on the reduced data using the same aperture selected for SOFI photometry (2.6$''$). 
For photometric calibration we used 30 2MASS sources with quality flag “A”. 


\subsubsection{Archival Photometry} 

We complemented our monitoring data with public domain photometry data. 
For optical bands we used Gaia G-band from the Gaia Photometric Science Alerts database, the Zwicky Transient Facility \citep[ZTF,][]{masci2019} DR17 $g'$- and $r'$-band from the ZTF archive and the Panoramic Survey Telescope \& Rapid Response System (PanSTaRRS). 

We used the ZTF data with “catflags = 0”, i.e., perfectly clean extracted, to filter out bad-quality images.
We downloaded the multi-epoch 2010-2014 Pan-STaRRS images obtained in $g'r'i'zY$ bands\footnote{\url{http://ps1images.stsci.edu/cgi-bin/ps1cutouts}}. 
We performed aperture photometry of Gaia18cjb using only the good quality images in 2.5$''$ aperture. Then, to calibrate the results to absolute magnitudes, we used the magnitude zero points provided in the headers for each image and filter. 
While this approach may result in increased light losses during poor seeing conditions, we estimated that it does not exceed 0.2~mag in the $g$-band, and $0.05-0.1$~mag for the other bands, which is sufficient for checking the light variability history before Gaia. As a consequence, we use 0.2\,dex of uncertainty for the $g'$-band photometry, and 0.1\,dex for the other bands photometry.
In addition, the smaller aperture increases the number of useful images, as their certain fraction suffers from detector imperfections. We compare our results to APASS9 $g'r'i'$ magnitudes of stars in the same field determined by \citet{henden2016} finding that their measurements are reproduced with our method within 0.05~mag.

For infrared wavelengths, we used $JHK_S$ photometric results from the Two Micron All Sky Survey \citep[2MASS,][]{cutri2003}, and downloaded processed $JHK_S$ images at two epochs from the UKIRT Infrared Deep Sky Survey \citep[UKIDSS,][]{lucas2008}. 
For UKIDSS images, in order to obtain results consistent with our SOFI and GTC infrared photometry results (Sect.~\ref{sec:sofi}), we performed aperture photometry using the same aperture and sky radii (in arcseconds) as for the SOFI and GTC data. 
Photometric calibration was performed via comparison with 15--25 2MASS sources, having quality flag ‘A’.
We also downloaded 3.35$\,\mu$m (W1) and 4.60$\,\mu$m (W2) photometry from the AllWISE Multiepoch Photometry Table and the NEOWISE-R Single Exposure (L1b) Source Table \citep{wright2010,mainzer2011,mainzer2014}, available at the NASA/IPAC Infrared Science Archive\footnote{\url{https://irsa.ipac.caltech.edu/Missions/wise.html}}. 
After filtering out bad photometry, e.g., due to smeared images, images affected by the South Atlantic Anomaly or by scattered light from the moon, we averaged the magnitudes taken in each observing season (WISE scans the sky and takes a few frames of a certain area once every 6 months). Saturation correction was not necessary. 


\subsection{Spectroscopy}
To probe the eruptive nature of Gaia18cjb, we observed it in the visible in one epoch and in the near-IR in three epochs, with different telescopes.
Technical details on the spectroscopic data are listed in Table\,\ref{tab:spectroscopy-text}. 

\subsubsection{LT SPRAT}  
\begin{figure}
    \centering
    \includegraphics[width=0.7\columnwidth]{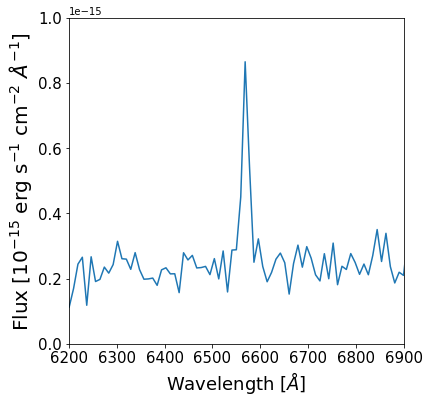}
    \caption{H$\alpha$ line of Gaia18cjb from LT SPRAT spectrum.}
    \label{fig:Halpha}
\end{figure}

On 2020 November 11 we observed Gaia18cjb in the visible at low resolution (R=350) with the SPectrograph for the Rapid Acquisition of Transients (SPRAT, \citealt{Piascik2014}), mounted on the Liverpool Telescope (LT)~\footnote{ProgID: XOL20B01, PI: Pawel Zielinski}. 
The spectrum was reduced and calibrated to absolute flux units by means of the dedicated SPRAT pipeline.
The low resolution and high noise of the spectrum prevent us from using it for any quantitative analysis, however, it clearly shows the presence of the H$\alpha$ line in emission (see Figure\,\ref{fig:Halpha}).

\subsubsection{LBT LUCI}
\begin{figure*}
    \centering
    \includegraphics[width=0.95\textwidth]{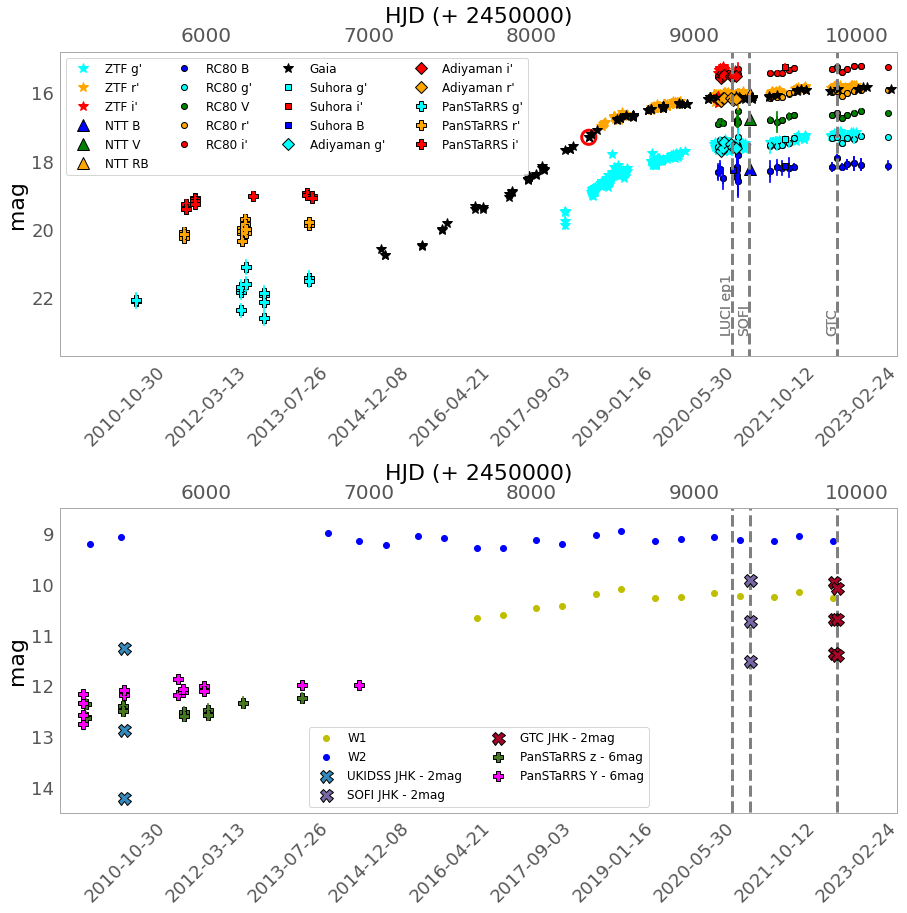}
    \caption{The visible ({\it top}) and  infrared ({\it bottom}) light curves of Gaia18cjb. The red circle highlights the Gaia Alert trigger. The error bars smaller than the symbol size are not presented. Grey dashed lines correspond to the epochs for which we collected NIR spectroscopy.}
    \label{fig:lightcurveVIS}
\end{figure*}

We observed Gaia18cjb at the Large Binocular Telescope (LBT Observatory, Arizona) with LUCI1 and LUCI2 intruments in the $zJ$ and $HK$ filters on 2021 January 17.  We reduced the data by using the SIPGI pipeline \citep{Gargiulo2022}, following the recipes indicated in the SIPGI documentation\footnote{\url{http://pandora.lambrate.inaf.it/docs/sipgi/}} for the flat field correction, sky subtraction, wavelength calibration, and telluric correction. 
We then averaged the LUCI1 and LUCI2 spectra, to maximize the signal-to-noise ratio (SNR). 
Since we do not have contemporary $zJHK$ photometry, we flux calibrate the spectrum using SOFI continuum as both optical and NIR light curves suggest that the photometry does not vary much (e.g. $\Delta m_r = 0.2$\,mag and $\Delta m_{W1} = 0.1$\,mag) between January and May 2021 (see Figure\,\ref{fig:lightcurveVIS}). 

\subsubsection{NTT SOFI} 
The blue part (0.95 -- 1.64$\,\mu$m) of NTT/SOFI spectrum was observed on the 2021 May 5 and 8, while the red part (1.53 -- 2.52$\,\mu$m) only on the 8th. 
For the observed data, flat-fielding, bad pixel removal, sky subtraction, aperture tracing, and wavelength calibration were applied with the Image Reduction and Analysis Facility \citep[IRAF,][]{Tody1993}. For the wavelength calibration, a Xenon spectrum was used. 
The hydrogen lines in the telluric standard stars were removed by Gaussian fitting. 
The spectra were flux calibrated using the contemporary photometry (see Sect.\,\ref{subsect:photometry}). 
Technical details are reported in Tab.\,\ref{tab:spectroscopy-text} and spectra are presented in Sect.\,\ref{sect:analysis}.

\subsubsection{GTC EMIR}
We obtained $JHK$-band spectroscopy of Gaia18cjb with the GTC/EMIR medium-resolution multi-object spectrograph \citep[Espectr\'ografo Multiobjeto Infra-Rojo,][]{Garzon2022} in long slit mode on 2022 October 26. 
EMIR is equipped with a 2048$\times$2048 Teledyne HAWAII-2 HgCdTe NIR-optimised chip with a pixel size of 0.2$\arcsec$. 
The total exposure time on source was 1920 seconds per grism. One grism per each band ($J$, $H$ and $K$) was used.  
A typical ABBA nodding pattern was applied. The seeing during the observations was FWHM$\sim$1.2$\arcsec$.

The spectra were reduced using several python routines customised by GTC staff for EMIR spectroscopic data. The sky background was first eliminated using consecutive A-B pairs. They were subsequently
flat-fielded, calibrated in wavelength and combined to obtain the final spectrum.
To correct for telluric absorption, we observed a telluric standard star with the same observing set up as the science target, right  after the Gaia18cjb observations and at similar airmass. To apply the correction we used a version of Xtellcor \citep{Vacca2003} specifically modified to account  for the atmospheric conditions of  La Palma observatory (\citealt{Ramos2009}). 
We calibrated the flux scale of GTC spectra using the
GTC contemporary photometry, see Sect.\,\ref{sec:sofi}.

\section{Results}
\label{sect:analysis}

Our photometric results, used for the optical and infrared light curves (see Sect.\,\ref{subsect:light curve}), are shown in Appendix\,\ref{app:photometry}. Uncertainties are computed as the quadratic sum of the uncertainty of the aperture photometry, the scatter of
the individual exposures, and the uncertainty of the photometric calibration.

Normalized and flux calibrated spectra are presented in Sect.\,\ref{sect:emission_lines} and Appendix\,\ref{app:spectra}, respectively. 
They show emission lines typical of YSOs such as magnetospheric accretion features (HI), winds/outflows (H$_2$), and CO/NaI from the disk. 

\subsection{Light curves}
\label{subsect:light curve}

Figure\,\ref{fig:lightcurveVIS} shows the optical (top panel) and IR (bottom panel) light curve of Gaia18cjb. 
\begin{figure*}
    \centering
    \includegraphics[width=1.5\columnwidth]{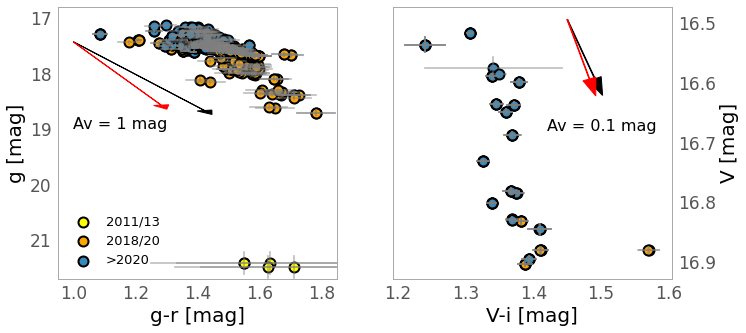}
    \includegraphics[width=1.5\columnwidth]{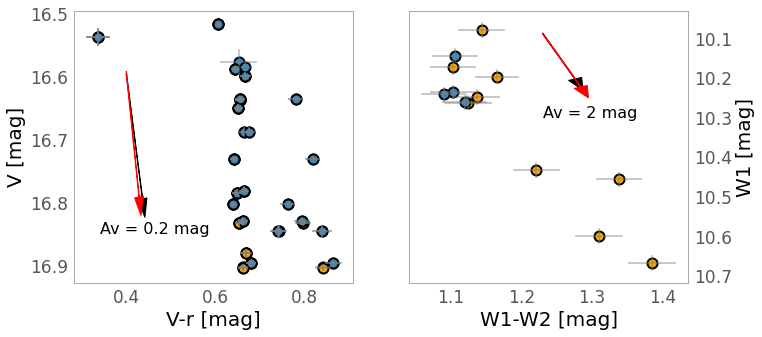}
    \caption{Color–magnitude diagrams of Gaia18cjb  based on data from PanSTaRRS data (yellow filled circles), ZTF, Konkoly, Mt Suhors, and WISE data before LUCI epoch (orange filled circles), and after LUCI epoch (blue filled circles). The arrows show the the reddening vector, 
    assuming extinction law from \citet{car89} with $R_V = 3.1$ (black arrow) and $R_V = 5.5$ (red arrow), respectively.}
    \label{fig:colordiagrams}
\end{figure*}

\subsubsection{Optical Light Curve}
The optical light curve shows the Gaia $G-$band brightening event of about 4.6\,mag from 2015 February to 2020 May (black stars). 
Afterwards, the brightness still increased, although slower, by only 0.3\,mag, up to the last epoch in March 2023. 
The overall brightening in the Gaia light curve is of about 4.9\,mag.
The outburst is still ongoing, and there is no indication that the peak of the outburst has been reached. However, it looks like Gaia18cjb has reached a constant plateau in 2022-23. 
The photometric data, collected from other observatories after 2016, cover timescales shorter than Gaia. 
They confirm the brightening and the current ``plateau" phase.
Indeed, the $r-$band data (orange stars, filled circles, triangles, diamonds, and crosses, depending on the telescope) show a strong increase of brightness consistent with the Gaia data acquired with the $G$ filter. 
The variability within 13 years is $\sim$5.0\,mag, a 
typical value for FUors variability \citep{fis22pp7}. However, the rise time (several years) is atypical for such kind of objects, although there are FUors-like objects whose light curve has been increased for decades \citep[e.g. V1515\,Cyg,][]{har96, Szabo2022}. 
The ZTF $g-$band shows brightening of 2.8\,mag within 4.75 years, and only of 0.5\,mag after 2021 January (LUCI Epoch).
NTT, RC80, Suhora, and ADYU60 light curves are in agreement within each other when the filter is the same and confirm the fact that since 2020 May the brightness of Gaia18cjb is still increasing but with a smaller amplitude than before (about 0.1\,mag) in the $i-$, $r-$, $V-$, $g-$, and $B-$ bands.

PanSTaRRS monitored Gaia18cjb prior to Gaia in $g-$, $r-$, and $i-$ bands from 2010 to 2013. 
During these years the variation in $r-$ band is only 0.65\,mag, suggesting a quiescent phase before the on-going outburst. 
The same behavior is also seen in other bands.
We note that the last data points from PanSTaRRs $r-$band are brighter than the first data points of Gaia, suggesting the presence of some possible pre-burst variability. 
We also note that PanSTaRRS $r$ and $g$ variability is larger in a certain epoch (May 2012) than in the overall three years. This can be a hint of ordinary small-amplitude accretion events as seen in CTTSs \citep[see e.g.][and references therein]{fis22pp7}.

\subsubsection{Infrared Light Curve}
We studied the variability also in the infrared using WISE, UKIDSS, and PanSTaRRS archival data together with our acquired photometry (bottom panel of Figure\,\ref{fig:lightcurveVIS}). 
The W1 channel at 3.4\,$\mu$m (yellow filled dots) varies by 0.6\,mag, while the variability is only 0.4\,mag at 4.6\,$\mu$m (W2 channel, blue dots) from 2010 March to 2022 October. 
The variability of both W1 and W2 channels reduced down to 0.1 since 2021 January. 
The W2 variability does not seem to follow the variability we see in the optical bands and it waves around a mean value between the pre-burst phase and the burst phase. 
We cannot confirm this trend in the W1 band since there were no W1 observations in the pre-burst phase. 
The IR light curve is similar to V1057\,Cyg, one of the classical FUors \citep{Szabo2021} after 1995 showing some kind of minor modulation.

The variability in the near-IR between UKIDSS (2007 April) and SOFI (2021 May) observations decreases with increasing wavelength: 2.70\,mag, 2.14\,mag, and 1.35\,mag in $J$, $H$, and $K$ bands, respectively. 
Interestingly, PanSTaRRS light curves, observed from 2010 to 2014, invert the trend: in the $Y-$band the variability is larger (0.89\,mag) than in the $z-$band (0.39\,mag). This can be explained by the large variability in the first epoch (2010 February), which can be due to the presence of a small burst or to a sudden increase of the light curve, as it is often the case before a strong FUor-like burst \citep{audard2014}. However, this could also be the result of the different sampling for the PanSTaRRS filters and, in turn, not being associated with the intrinsic variability of the star.

We highlight that the photometric variability in the infrared is moderate in general and it does not reflect the strong burst we observe in the visible. 
Indeed, the infrared variability is as small as in typical non eruptive sources \citep[see, for example,][]{lorenzetti2012, zsidi2022vw,fiorellino2022wx}. 
However, this result is puzzling since it suggests that the burst is restricted to the star and its immediate circumstellar environment and does not affect the dusty disk at all. We will discuss possible interpretations of this result in Sect.\,\ref{sect:discussion}.

\subsection{Color-magnitude diagrams}
\label{sect:colcoldiagrams}

To investigate the photometric properties of Gaia18cjb in the visible and infrared bands, we built color–magnitude diagrams which are in Figure\,\ref{fig:colordiagrams}. 
The top-left panel corresponds to the $g'$ versus $g'-r'$ diagram. 
It shows extinction related to color variations, and a linear decreasing trend.
This trend suggests that $g'$- and $r'$-band variations are correlated, but $g'$-band variation is larger than $r'$-band variations (see light curves on Figure\,\ref{fig:lightcurveVIS}). 
A similar trend is also present in the infrared color-magnitude diagram in the bottom-right panel, where W1 and W1$-$W2 strongly correlate following the direction of the reddening vectors.  
It appears that the source is redder when fainter, hence the downward trend in the color–magnitude diagram. 
The top-right panel and the bottom-left panel present the $V$-band photometry versus $V-i'$ and $V-r'$, respectively. 
In both cases, the distribution appears to be vertical, suggesting that the variability amplitude is similar in the $V$ and $i'$ and in the $V$ and $r'$ bands.
In all the color-magnitude diagrams in Figure\,\ref{fig:colordiagrams} we note a temporal trend: the variation of the colors decreases with time as the source brightens. 
This produces a trend on the opposite direction of the reddending vectors. 
In general, the difference between the reddening law with $R_V = 3.1$ (black arrow) and $R_V = 5.5$ (red arrow) is small, and it is hard to distinguish which one of the two is followed better from the data points. 
In the $g$\,vs.\,$g-r$ plot this difference is more evident and the black arrow better reproduces the variation of extinction from 2018 to 2020 (orange-blue filled dots), while the steeper slope of the red arrow better reproduces $A_V$ variations in the quiescent phase and in the latest epochs (from 2021 onward, yellow-blue filled dots). 
We discuss possible interpretations of this result in Sect.\,\ref{sect:discussion}.

\begin{figure}
    \centering
    \includegraphics[width=0.7\columnwidth]{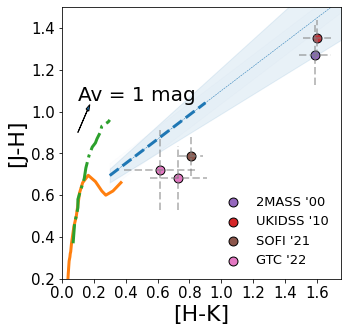}
    \caption{NIR color–color diagram of Gaia18cjb. Purple and red circles are pre-outburst data from 2MASS and UKIDSS observations in 2000 and 2010, respectively, while the brown and pink dots correspond to SOFI and GTC photometry in the burst phase (2021 and 2022, respectively). 
The blue dashed line and light blue region represent the
CTTS locus and its uncertainty \citep{mey97}, and the tiny blue dotted line its extension toward larger color indices. The orange solid and green dotted–dashed lines correspond to the colors of zero-age main sequence and giant branch stars, respectively. The black arrow shows the reddening vector for a source with 1 mag of extinction \citep[][$R_V = 3.1$]{car89}.
}
    \label{fig:colorcolordiagram}
\end{figure}

Figure\,\ref{fig:colorcolordiagram} shows the $J-H$\,vs.\,$H-K$ diagram, which can be used to classify a YSO and to estimate its extinction. 
We used UKIDSS and 2MASS archival data and the newly acquired SOFI and GTC data points. 
We also plot in the same figure the giant branch locus (green dot-dashed line), the Zero Age Main Sequence (ZAMS, orange line), and the Classical T\,Tauri Stars (CTTS) locus \citep{mey97}, where CTTS are not extincted ($A_V = 0 $\,mag, blue dashed line).
By  following the extinction vector by \citet{car89} and $R_V = 3.1$ (black arrow), it is possible to  compute the extinction of a CTTS.
The 2MASS and UKIDSS data of Gaia18cjb lie within the CTTS locus, suggesting negligible extinction, while recent data points lie below the CTTS locus, making not possible to compute its extinction with this method and suggesting that Gaia18cjb is less evolved than a typical CTTS. 
However, in the case of GTC data, considering the error bars, it is possible that the Gaia18cjb colors lie on the CTTS locus, suggesting low extinction. 
This movement along the CTTS locus from redder to bluer is similar to what was found in \citet[][see Figure\,7 therein]{kospal2011} for HBC\,722.
In general, it is known that highly accreting stars become bluer in burst with respect to quiescence \citep[see e.g.][]{lorenzetti2007, lorenzetti2011, hillenbrand2018, Szegedi-Elek2020}, suggesting the accreting nature of Gaia18cjb. 

\subsection{Spectral Energy Distribution}
\label{sect:SED}
\begin{figure}
    \centering
    \includegraphics[width=\columnwidth]{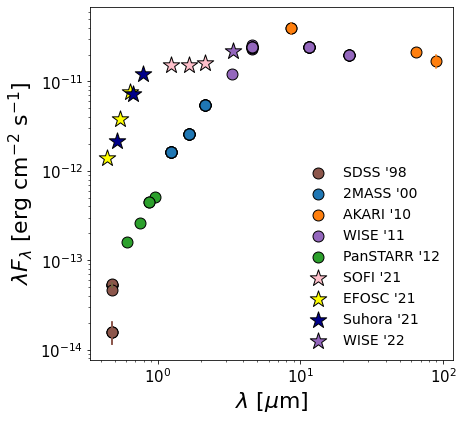}
    \caption{Spectral Energy Distribution. Circles and stars are observations taken before and after 2020, respectively.}
    \label{fig:sed}
\end{figure}

From an observational point of view, YSOs are usually classified on the basis of their spectral energy distribution \citep[SED,][]{lad84,lad87,and95,Greene1994}.
Figure~\ref{fig:sed} shows the SED of Gaia18cjb, in the low brightening state (dots, before 2020), resembling the SED of an embedded object as a Class\,I YSOs, and in the outburst state (after 2020, star symbols), which is similar to a Flat Spectrum source. 
The data we used to build up the SEDs were observed at different periods and are listed in  Appendix\,\ref{app:photometry}. 

Note that, based on the light curve, it is challenging to determine when the outburst begun. 
The Gaia alert was triggered in 2018, but the luminosity had already increased since 2015 (when first Gaia data are available), while the PanSTARRS monitoring between 2010 and 2014 does not provide definitive conclusions about the {\it quiescent} phase of this object, even if we can tentatively assume the average PanSTARRS magnitudes as values for the quiescent phase ($m_g = 21.8 \pm 0.4$, $m_r = 20.0 \pm 0.2$, $m_i=19.1 \pm 0.1$, $m_z = 18.5 \pm 0.1$, and $m_Y = 18.2 \pm 0.2$).
However, we distinguish here between the first part of the optical light curve, where the luminosity increases by 4.6\,mag in about 10 years, and the second part, after LUCI epoch, where the luminosity moderately increases by only 0.3\,mag in three years (see the top panel of Figure\,\ref{fig:lightcurveVIS} and Sect.\,\ref{subsect:light curve}). 
We refer to the first one as the {\it brightening} or {\it low state} phase and to the latter one as the {\it bursting} or {\it high state} phase. 
We stress that this nomenclature is only useful to distinguish between the amplitude in the variability while in both the phases the burst is on, with the luminosity increasing.

We computed the spectral index $\alpha = \frac{d\log (F_\lambda \lambda)}{d\log (\lambda)}$ between 2.16\,$\mu$m and 22.1\,$\mu$m using 2MASS and WISE data, respectively. 
For the pre-outburst phase, we obtain $\alpha = 0.55 \pm 0.28$, which classifies Gaia18cjb as a Class\,I protostar ($\alpha > 0.3$). We note that the spectral index is also compatible with a Flat Spectrum object ($-0.3 < \alpha < 0.3$) within the error. We conclude Gaia18cjb is most likely a Class\,I YSOs, with a small chance that it is a Flat Spectrum. 
We also note that the classification we use computes the spectral index using $\lambda F_\lambda$ at 2\,$\mu$m and 24\,$\mu$m (not 22\,$\mu$m as we do) but, based on the SED shape, we do not expect a large difference between the flux at 22\,$\mu$m and at 24\,$\mu$m.  

We also computed the spectral index using SOFI $K-$band and WISE W4 data, to get an estimate of $\alpha$ for the eruptive phase. 
SOFI observations were taken during the eruptive phase, while WISE observations are not, but since the low variability at long wavelengths and the lack of observations around 20\,$\mu$m during the outburst, we use the same value as we did considering the brightening SED. 
We obtain $\alpha = 0.09 \pm 0.02$, compatible with a Flat Spectrum.

In both the low level and high level phases, the spectral index we compute does not describe Gaia18cjb as a Class\,II PMS young star.  
In the scenario where the empirical classification of YSOs based on the SED spectral index corresponds to an evolutionary classification, Gaia18cjb seems to be in between the protostellar phase, where the star is embedded in its envelope and the accretion is very high (Class 0 and I YSOs), and the pre-main sequence phase, where most of the stellar mass is accessed and the star is approaching the ZAMS (Class II and III YSOs).
Estimates of Class\,0 lifetimes are between 0.2 and 0.6\,Myr, while, based on {\it Spitzer} timescales, lifetimes of Class\,I and Flat\,Spectrum YSOs are 0.54\,Myr and 0.40\,Myr, respectively \citep{eva09}. Thus, if the classification based on SED spectral index corresponds to evolutionary stages and our classification is correct, we can assume for Gaia18cjb an age of about 1\,Myr. 

\subsection{Emission Lines}
\label{sect:emission_lines}

\begin{figure*}
    \centering
    \includegraphics[width=\columnwidth]{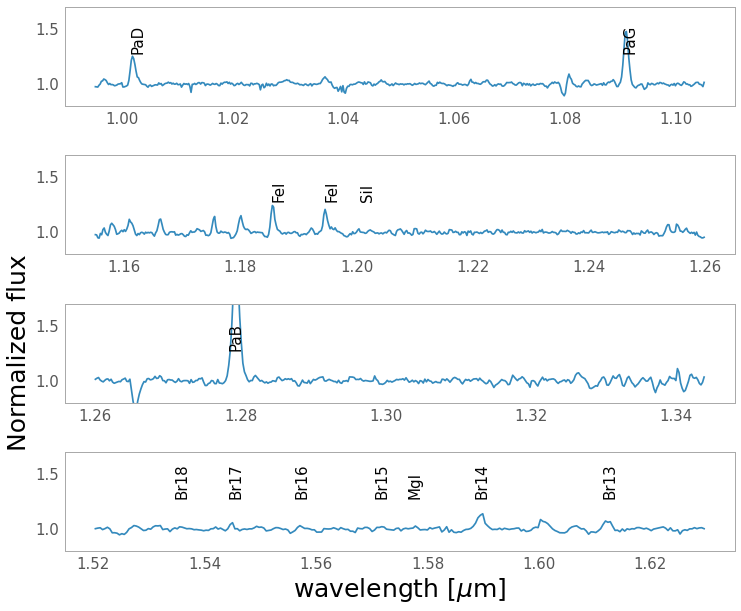}
    \includegraphics[width=\columnwidth]{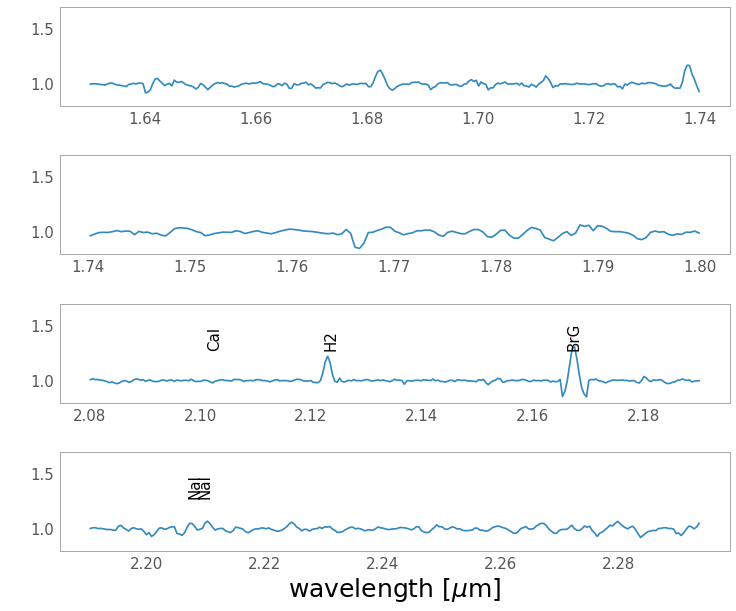}
    
    \caption{Normalized LBT zJHK normalized spectrum. Emission lines reported in Table\,\ref{tab:EW} are labelled.}
    \label{fig:LBTspec}
\end{figure*}

\begin{figure*}
    \centering
    \includegraphics[width=\columnwidth]{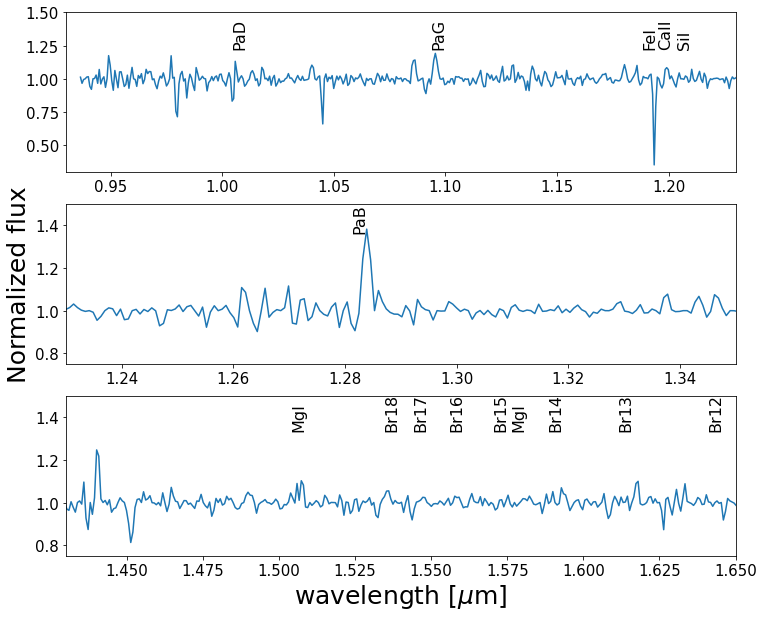}
    \includegraphics[width=0.96\columnwidth]{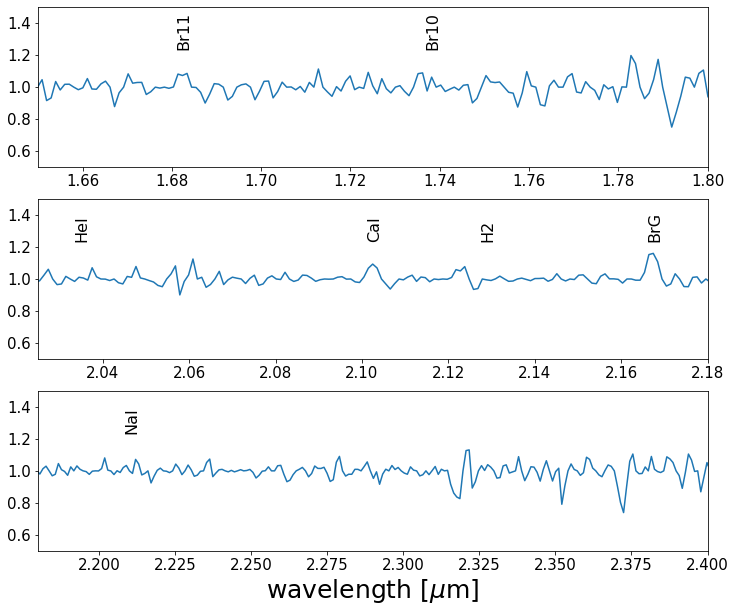}
    
    \caption{Normalized SOFI spectrum. Emission lines reported in Table\,\ref{tab:EW} are labelled.}
    \label{fig:SOFIspec}
\end{figure*}

\begin{figure*}
    \centering
    \includegraphics[width=\columnwidth]{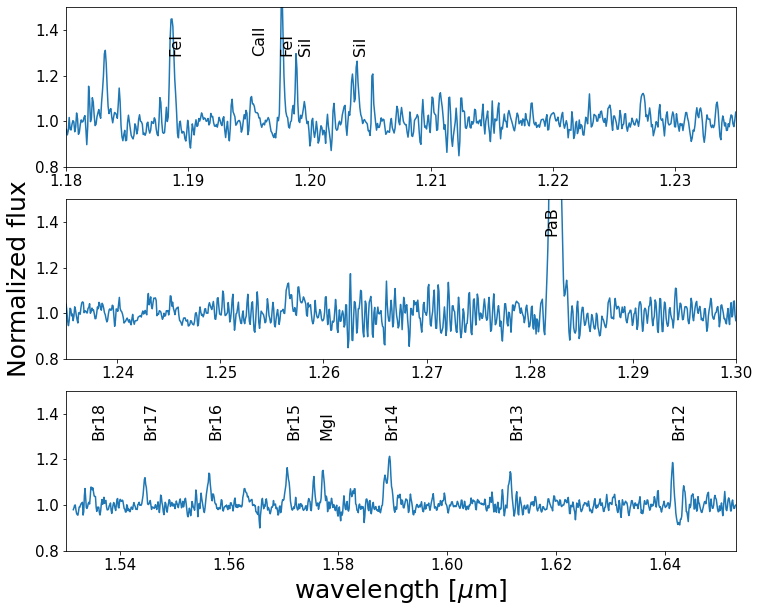}
    \includegraphics[width=\columnwidth]{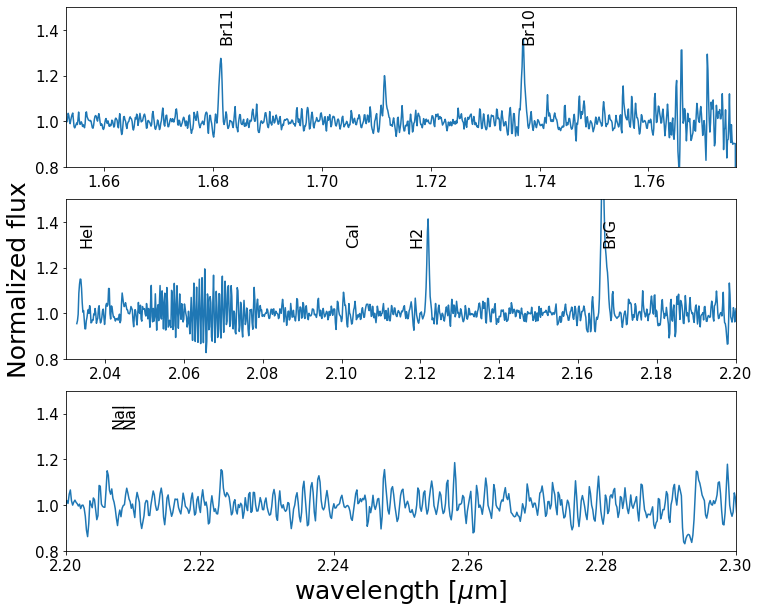}

    \caption{Normalized GTC JHK spectrum. Emission lines reported in Table\,\ref{tab:EW} are labelled.}
    \label{fig:GTCspec}
\end{figure*}

\begin{table}
\centering
\caption{\label{tab:EW}Equivalent width of Gaia18cjb emission lines for the three epochs we observed. The equivalent width unit is \AA.}
\begin{tabular}{lrccc}
\hline
\hline
Line & $\lambda$ & W$_{eq}^{\rm LUCI}$ & W$_{eq}^{\rm SOFI}$ & W$_{eq}^{\rm GTC}$\\

& [$\mu$m] & \\
\hline		
\hline
Pa$\beta$        & 1.281 & 18.41$\pm$6.22 & 14.62$\pm$1.66 & 21.99$\pm$3.01\\		
${\rm Pa\gamma}$ & 1.093 & 5.99$\pm$0.96  &  7.20$\pm$1.98 & $-$ \\                  
${\rm Pa\delta}$ & 1.004 & 4.58$\pm$0.97  & $<$2.01        & $-$ \\                  
${\rm Br\gamma}$ & 2.166 & 11.90$\pm$1.36 & 10.49$\pm$3.08 & 10.51$\pm$0.88\\                     
Br10             & 1.736 & 3.89$\pm$0.66  & $<3.93$        & 3.12$\pm$0.7  \\                   
Br11             & 1.681 & 4.1$\pm$1.12   & $<6.99$        & 2.06$\pm$0.56  \\                   
Br12             & 1.641 & 4.83$\pm$1.33  & $<4.77$        &  1.61$\pm$0.2\\                     
Br13             & 1.611 & 2.17$\pm$0.56  & $<8.94$        & 1.14$\pm$0.43 \\                    
Br14             & 1.588 & 3.20$\pm$1.12  & $<1.60$        & 2.77$\pm$0.31 \\                     
Br15             & 1.570 & $<2.90$        & $<2.07$        & 1.08$\pm$0.12 \\                     
Br16             & 1.556 & $<3.81$        & 1.16$\pm$0.37  & 1.41$\pm$0.16 \\                    
Br17             & 1.544 & $<2.28$        & $<2.31$        & 1.04$\pm$0.27 \\                     
Br18             & 1.534 & $<4.87$        & 1.80$\pm$0.42  & $<$2.73 \\                     
Fe I             & 1.188 & 1.02$\pm$0.15  & $<5.39$        & 2.3$\pm$0.36 \\        
Fe I             & 1.197 & 2.84$\pm$0.97  & $<8.05$        & 2.06$\pm$0.65\\        
Ca II            & 1.195 & $<2.76$        & $-$            & $<$1.05 \\              
Si I             & 1.203 & $<0.67$        & $<7.39$        & 1.71$\pm$0.64\\        
Mg I             & 1.502 & $-$            & $<1.61$        & $-$ \\          
Mg I             & 1.504 & $-$            & 3.45$\pm$1.11  & $-$\\                  
Mg I             & 1.576 & $<2.72$        & $<10.94$       & 0.96$\pm$0.18\\        
He I             & 2.033 & $-$            & $<0.85$        & 1.73$\pm$0.51\\        
Ca I             & 2.101 & $<1.35$        & 4.20$\pm$1.10  & 0.57$\pm$0.22 \\
Na I             & 2.206 & 2.08$\pm$0.43  & 2.21$\pm$0.49  & $<$1.35 \\
Na I             & 2.208 & 2.18$\pm$0.40  & 2.21$\pm$0.49  & 0.92$\pm$0.32\\
H$_2$            & 2.127 & 3.59$\pm$0.71  & 2.71$\pm$0.57  & 3.57$\pm$0.45\\
\hline
\hline
\end{tabular}
 \end{table}

The spectra of young solar-type stars show many features tracing the star-disk interaction as, for example, accretion/ejection.  
Our NIR spectra of Gaia18cjb covers several emission lines, and in particular some accretion tracers, such as $\brg$ and $\pab$. 
We describe here how we have measured the fluxes of these lines to be used for the analysis.

We measured the equivalent width ($W_{eq}$) and the observed line flux ($F_{obs}$) from the normalized and flux calibrated spectra, respectively, as follows. We fitted a linear curve to the local continuum in a wavelength range of $\Delta \lambda = 2\,nm$, centered on the emission line wavelength $\lambda_0$. 
We slightly modified this range, if needed, taking the one most suitable for each line; for example, by avoiding other emission lines, if present, or telluric absorption lines. The line flux was determined by subtracting the local continuum from the spectra and integrating it over the line. We computed the noise of the line by multiplying the standard deviation of the local continuum ($rms$) for the wavelength element between two pixels $\Delta \lambda$, and multiplying this by the square root of the number of pixels within the wavelength range ($N_{pix}$). We considered a line to be detected when its $S/N > 3$. For those lines that were detected in at least one epoch, we estimated the upper limits in the other epochs as three times the noise: 
\begin{equation}
    F_{line}^{upp} = 3 \times (\sqrt{N_{pix}} \times rms \times \Delta \lambda)
\end{equation}
For LUCI spectra, the flux calibration with contemporary photometry was not possible. Therefore, we computed the observed flux by multiplying the equivalent width of each line by its local continuum in the SOFI spectrum. 
We stress that the acquisition of the LUCI spectrum was contemporaneous to that of the SOFI data. 
Equivalent widths and fluxes are reported in Tabs.\,\ref{tab:EW} and \ref{tab:flux_obs}, respectively.

{\it LBT spectrum: 2021 January.} Figure\,\ref{fig:LBTspec} reports the LBT normalized spectrum and the emission lines detected. 
This spectrum shows several accretion tracers such as Pa$\beta$, Pa$\gamma$, Pa$\delta$, Br$\gamma$, Br10, Br11, Br12, Br13, Br14, and the Na\,I doublet. 
The H$_2$ line in emission suggests the presence of some shocked material in ejection, as winds. We also detect Fe\,I in emission.

{\it SOFI spectrum: 2021 May.} Figure\,\ref{fig:SOFIspec} shows the SOFI normalized spectrum. It confirms the presence of accretion (Pa$\beta$, Pa$\gamma$, Br$\gamma$, Br16, Br18, Na\,I doublet) and ejection (H$_2$) tracers. Emission lines of Mg\,I and Ca\,I are also detected. 

{\it GTC spectrum: 2022 October.} The GTC spectrum shows up to 9 lines of the Brackett series, but, due to the restricted wavelength range in the blue part of the spectrum, only the brightest line of the Paschen series (Pa$\beta$ see Figure\,\ref{fig:GTCspec}). Fe\,I, Mg\,I, Ca\,I, and Na\,I lines are detected. 
The H$_2$ emission is confirmed. 

\begin{table}
\centering
\caption{\label{tab:flux_obs}Observed fluxes of the emission lines in the Gaia18cjb spectra. The observed flux unit is $10^{-14}$ erg s$^{-1}$ cm$^{-2}$.}
\begin{tabular}{lrccc}
\hline
\hline
Line & $\lambda$ & F$_{obs}^{\rm LUCI}$ & F$_{obs}^{\rm SOFI}$ & F$_{obs}^{\rm GTC}$\\

 & [$\mu$m] & \\
\hline		
\hline
Pa$\beta$        & 1.281 & 1.08$\pm$0.07  & 1.96$\pm$0.16 & 0.664$\pm$0.046\\								
${\rm Pa\gamma}$ & 1.093 & 0.73$\pm$0.12  & 0.72$\pm$0.10 & $-$ \\                          
${\rm Pa\delta}$ & 1.004 & 0.57$\pm$0.12  & $<$0.25       &  $-$ \\                         
${\rm Br\gamma}$ & 2.166 & 0.88$\pm$0.10  & 1.05$\pm$0.11 & 0.680$\pm$0.045\\               
Br10             & 1.736 & 0.30$\pm$0.07  & $<$0.25       & 0.078$\pm$0.016 \\                             
Br11             & 1.681 & 0.60$\pm$0.08  & $<$0.29       & 0.051$\pm$0.013\\                              
Br12             & 1.641 & 0.49$\pm$0.18  & $<$0.29       & 0.041$\pm$0.005\\                               
Br13             & 1.611 & 0.21$\pm$0.05  & $<$0.43       & 0.029$\pm$0.01\\         
Br14             & 1.588 & 0.32$\pm$0.11  & $<$0.40       & 0.072$\pm$0.007\\         
Br15             & 1.570 & $<$            & $-$           & 0.029$\pm$0.003\\              
Br16             & 1.556 & $<$0.36        & 0.12$\pm$0.04 & 0.037$\pm$0.004\\                             
Br17             & 1.544 & $<$0.37        & $<$0.33       & 0.028$\pm$0.007\\                             
Br18             & 1.534 & $<$0.37        & 0.20$\pm$0.05 & $<$0.07\\                              
Fe I             & 1.188 & 0.08$\pm$0.01  & $<$0.14       & 0.074$\pm$0.01\\                     
Fe I             & 1.197 & 0.35$\pm$0.12  & $<$0.84       & 0.068$\pm$0.018\\                       
Ca II            & 1.195 & $<$1.02        & $<$0.80       & 0.032\\                              
Si I             & 1.203 & $<$0.25        & $<$0.80       & 0.055$\pm$0.019\\                              
Mg I             & 1.502 & $-$            & $<$0.56       & $-$ \\                          
Mg I             & 1.504 & $-$            & 0.39$\pm$0.13 & $-$ \\                          
Mg I             & 1.576 & $<$0.85        & $<$3.44       & 0.026$\pm$0.005\\
He I             & 2.033 & $-$            & $<$0.54       & 0.118$\pm$0.032\\   
Ca I             & 2.101 & $<$0.27        & 0.24$\pm$0.05 & 0.037$\pm$0.040\\
Na I             & 2.206 & 0.18$\pm$0.04  & 0.19$\pm$0.04 & $<$0.087\\                       
Na I             & 2.208 & 0.18$\pm$0.03  & 0.19$\pm$0.04 & 0.062$\pm$0.021\\
H$_2$            & 2.127 & 0.25$\pm$0.05  & 0.67$\pm$0.18 & 0.232$\pm$0.026\\
\hline
\hline
\end{tabular}
\begin{quotation}
  \textbf{Notes.} Observed fluxes are not corrected for the extinction. GTC and SOFI lines flux is computed from the flux calibrated spectra, while LUCI lines flux is computed from the $W_{eq}$ using the SOFI continuum for that line. 
  \end{quotation} 
\end{table}

\section{Discussion}
\label{sect:discussion}

In this section we discuss evidences supporting that Gaia18cjb  is a YSO, and put into context its accretion properties with respect to other erupting YSOs. 

\subsection{On the distance of Gaia18cjb}
\label{App:distance}
Measuring the distance of an unknown object is a complex task. In particular, the case of Gaia18cjb presents its own set of challenges, given the large parallax uncertainty derived by Gaia. 
In this session we discuss strenghts and limitations of the photogeometric distance by \citet{bai21}.

The Gaia DR3 parallax for Gaia18cjb is $p = 0.83 \pm 0.69$\,mas. 
With such large uncertainty, the parallax method does not yield accurate results.
Indeed, the main DR3 catalog does not provide a distance for sources whose parallax has low SNR, including our target.
For this reason, \citet{bai21} developed a probabilistic approach to estimate stellar distances. 
This method relies on the prior using a full range of populations in the mock catalogue, based on the Besancon Galaxy model.
For each source they computed two posterior probability distributions over distance: a geometric distance, based only on the prior and the parallax, and the photogeometric distance, based also on the G magnitude and the BP$-$RP color. 
Gaia18cjb's geometrical distance is $d_{g} = 3.20_{1.62}^{2.66}$\,kpc, while the photogeometrical distance is $d_{pg}=1.03^{+0.20}_{-0.19}$\,kpc. 
The recommendation of \citet{bai21} is to adopt the photogeometric distance for sources whose parallax has low SNR. This is the case of Gaia18cjb.

The point of the distance inference in \citet{bai21} is to provide a gradual transition from data-dominated to prior-dominated distance estimates. 
While the populations used by \citet{bai21} contained YSOs, their results are not optimized for intrinsically red sources with high circumstellar extinction, such as YSOs.
While a more appropriate photometric prior constructed solely for young stars would provide more accurate photometric distances, this is not the primary objective of our study.
To be conservative, we here discuss how our main results change, considering the spread in the distance suggested from Gaia parallax $d \sim 1/p = 660\,\mbox{pc}\, - 7140$\,pc.

In the following, we discuss results assuming $d_{min} = 660$\,pc, $d_{BJ} = 1030$\,pc, and $d_{max} = 7140$\,pc.

\begin{table*}
\centering
\caption{\label{tab:Av_int}Interstellar extinction and bolometric luminosity at different distances.}
\begin{tabular}{rccccccc}
\hline
\hline
Distance & A$_V^{B19}$ & A$_V^{STILIsM}$ &  A$_V^{mean}$ & $\lbol^{pre}$ & $\lbol^{pre\_corr}$ & $\lbol^{burst}$ & $\lbol^{burst\_corr}$ \\
pc & mag & mag & mag & $\lsun$ & $\lsun$ & $\lsun$ & $\lsun$\\ 
\hline		
\hline
660 & $0.3 \pm 0.1$ & $0.5 \pm 0.2$ & $0.4 \pm 0.2$ & $1.4\pm0.2$ & $1.4\pm0.2$ & $3.0\pm0.6$ & $3.1 \pm 0.6$\\
1030 & $0.8 \pm 0.2$ & $1.0 \pm 0.2$ & $0.9 \pm 0.3$ & $3.4 \pm 0.2$ & $3.5\pm0.2$ & $8.0 \pm 0.6$ & $8.3 \pm 0.6$\\
7140 & $2.9 \pm 0.4$ & $ 3.1\pm 0.4$ & $ 3.0\pm 0.6$ & $162 \pm 31$ & $173 \pm 31$ & $353 \pm 31$ & $534 \pm 31$\\
\hline
\hline
\end{tabular}
 \end{table*}

\subsubsection{Interstellar Extinction}
\label{sect:extinction}

To compute the reddening of Gaia18cjb due to the interstellar extinction, we use the 3D dust maps based on Gaia parallaxes and stellar photometry from Pan-STaRRS\,1 and 2MASS \citep{green_2019}. We run the \url{dustmaps} python package\footnote{\url{http://argonaut.skymaps.info/usage}}, using the most updated version {\it Bayestar 2019}, assuming $R_V = 3.1$, and the reddening law by \citet{car89}. 
We performed the same exercise using the STructuring by Inversion the Local Interstellar Medium \citep[STILIsM,][]{Capitanio2017, Lallement2018} with the on-line tool\footnote{https://stilism.obspm.fr/}. 
In both cases, we consider these values as lower limits for the extinction towards the source, because they are values averaged in a certain area and, most importantly, not sensitive to local circumstellar extinction.
We repeate the calculation for $d_{min}$, $d_{BJ}$, and $d_{max}$. 
In Tab.\,\ref{tab:Av_int} $A_V$ results are listed, ranging from 0.4\,mag to 3.0\,mag depending on the distance. We note that extinction estimates obtained with STILIsM show slightly higher values, but compatible within the errors.

 \subsubsection{Bolometric Luminosity} \label{sect_Lbol}
Using the fluxes reported in Appendix.\,\ref{app:photometry}, we computed the bolometric luminosity by integrating the SED fluxes ($F_\lambda$) with a dedicated python procedure, following the same method already used in literature 
\citep[e.g.][]{ant08, fio21, Fiorellino2023}. 
We integrated the SED from the shortest wavelength, interpolating with straight lines in the $\log \lambda - \log F_\lambda$ plane between the available SED points. 
We denote the value found with this method $\lbol^{pre}$, to highlight that this is the bolometric luminosity corresponding to the pre-burst phase, obtained using the photometry observed before  2015 (filled circles in Fig.\,\ref{fig:sed}).

We added optical and NIR photometry acquired during the outburst (i.e. after 2015, filled stars). 
For the high level phase, we computed the bolometric luminosity using Suhora, NTT/EFOSC, NTT/SOFI, and W1 photometry (blue, yellow, pink, and purple stars in the figure) instead of SDSS, PanSTARR, and 2MASS photometry (brown, green, and blue filled circles in the figure). 
As there is little or no variability in W2, we assumed that the source was constant within about 0.4\,mag at 4.5\,$\mu$m, and used the ALLWISE and AKARI points reward these wavelengths for the outburst SED, too. 
We name the value found with this method $\lbol^{burst}$, to highlight that this is the bolometric luminosity computed using the photometry observed during the ourtburst (see Fig.\,\ref{fig:sed}).

To build the SED and to measure the bolometric luminosity of Gaia18cjb we used observed fluxes, not corrected for the extinction.
We note that the longest wavelength data point we have is AKARI at 90\,$\mu$m, and we assume a decrease in the emission as $1/\lambda^2$ after that. Even applying this correction, we are aware that the bolometric luminosity derived so can be underestimated due to the lack of information at longer wavelengths. 
We also assume that beyond 3.6\,$\mu$m the SED did not change during the burst.

It is desirable to correct the flux for extinction and compute the bolometric luminosity corrected for the extinction ($L_{bol}^{corr}$). Following \citet{eva09},  
since the spectral index suggest Gaia18cjb can be classified as a Class\,I/Flat Spectrum YSO, we used the interstellar $A_V$ to remove foreground extinction, but not local extinction from the surrounding envelope, which will be reradiated in the far-infrared. We used the mean value between the values we obtain with the \url{dustmaps} and STILIsM softwares. Results are listed in Tab.\,\ref{tab:Av_int}.

The lowest possible distance and the BJ distance both results in pre-outburst luminosities ($1.4 - 3.5 \lsun$) typical of a low-mass young protostar, and the outburst luminosities ($3.1 - 8.3 \lsun$) are reasonable for modest accretion outbursts (cf. HBC 722 or some EXors).
The luminosities obtained with the largest possible distance are in the several hundred $\lsun$ range both in quiescence ($173 \lsun$) and in outburst ($534 \lsun$). While such a large outburst luminosity is possible in some of the most luminous and most highly accreting FUors \citep[e.g., V1057 Cyg or FU Ori][and refs therein]{audard2014}, the quiescence luminosity is definitely too large for a low-mass YSO.

\subsubsection{Stellar parameters and Accretion Rates}
\label{sect:stellar_luminosity}

In this section, the stellar parameters (spectral type, extinction, luminosity, radius and mass) and accretion properties (accretion luminosity and accretion rate) of Gaia18cjb are determined.

\begin{figure}
    \centering
    \includegraphics[width=0.9\columnwidth]{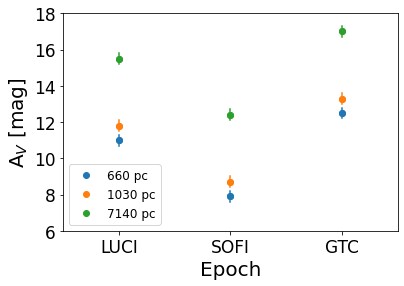}
    \caption{Extinction estimates of Gaia18cjb computed from the empirical relations between the accretion luminosity and Br$\gamma$ and Pa$\beta$ lines. Different colors represent $A_V$ estimates at different distances as described in the legend. 
    }
    \label{fig:Av}
\end{figure}

An estimate of the total extinction can be done by assuming  that the accretion luminosity, $\lacc$, as derived from different accretion tracers, is the same. 
In order to derive $\lacc$, we used the Pa$\beta$ and Br$\gamma$ lines and the empirical relationships between $\lacc$ and the line luminosity, $L_{\rm line}$, by \citet{alc17}.  By  minimizing the quantity 
$|\lacc(Pa\beta)-\lacc(Br\gamma)|$ it is possible to estimate the circumstellar extinction, $A_V$, as follows: the observed fluxes of the Pa$\beta$ and Br$\gamma$ lines were dereddened with a series of $A_V$ values ranging from 0 to 200\,mag in steps of 0.1\,mag. 
The $A_V$ values minimizing the $|\lacc(Pa\beta)-\lacc(Br\gamma)|$ quantity are given in Table\,\ref{tab:AvLacc} and shown graphically in Figure\,\ref{fig:Av}. 
For these, we determine values of $\log(\lacc/\lsun)$ between $-0.07$ and 2.63\,dex, depending on the chosen distance (see Tab.\,\ref{tab:AvLacc}). 
A robust value of $A_V$ for the eruptive phase can be then obtained by computing the average from the three individual estimates, the standard deviation being the error. This value ranges from $A_V = 10.5 \pm 2.7$\,mag to $A_V = 15.0 \pm 0.3$\,mag, depending on the distance, see Tab.\,\ref{tab:AvLacc}.

Assuming that $\lbol = \lstar + \lacc$, we get a rough estimate of the stellar luminosity for Gaia18cjb by knowing the dereddened bolometric luminosity for the eruptive phase $\lbol^{burst\_corr}$ and the accretion luminosity $\lacc$. 
We thus obtain $\lstar = 1.9 \pm 1.2 \, \lsun$ for a $d_{min} = 660$\,pc, $\lstar = 6.0 \pm 1.2 \, \lsun$ for $d_{BJ} = 1030$\,pc, and $\lstar = 107 \pm 62 \, \lsun$ for $d_{max} =7140$\,pc.

Based on the spectral features, we discard the possibility for Gaia18cjb to be a high mass star, since high excitation lines, such as He\,I are not detected.
The stellar luminosity we obtain if assuming a distance of 7140\,pc ($\lstar = 107 \pm 62 \, \lsun$) is, therefore, not compatible with information contained in our spectra.
Our spectra also discard the possibility that the spectral type of Gaia18cjb could be later than K7. Indeed, M-type and later type sperctra show photospheric absorption features such as CO and VO bands. 
One may argue that the molecular bands could be veiled by the strong continuum emission, but if that were the case then the accretion luminosity should be extremely high, and the W$_{eq}$ we obtain suggest this is not the case.

Then, we used the evolutionary tracks by \citet{sie00}, assuming an age of $\sim 1$\,Myr (see Sect.\,\ref{sect:SED}) and considering only spectral type earlier than K7. 
We found that our results for $\lstar$ at 660\,pc are compatible with a young star with stellar mass $\mstar = (0.7 - 1.6) \, \msun$, a radius $\rstar = (3.5 - 1.4)\, \rsun$, and $\teff = 4000 - 4600$\,K, corresponding to SpT$=$K7-K3.
The same exercise for $\lstar$ at 1030\,pc leads to a young star with stellar mass $\mstar = (0.7 - 2.2) \, \msun$, a radius $\rstar = (4.4 - 2.6)\, \rsun$, and $\teff = 4000 - 5080$\,K, corresponding to SpT$=$K1-K7.

From the derived $\lacc$ and stellar parameters, we computed the mass accretion rate ($\macc$) in the bursting phase using the relation from \citet{har98} in the framework of magnetospheric accretion scenario:
\begin{equation}
\macc = \Bigg( 1 - \frac{\rstar}{R_{\rm in}} \Bigg)^{-1} 
\frac{\lacc \rstar}{G \mstar}
\label{eq:macc}
\end{equation}
where $R_{in}$ is the inner-disk radius which we assume to be $R_{in} \sim 5\rstar$ as discussed by \citet[][see their Eq.\,3]{har16} and for consistency with previous similar studies \citep[e.g.][]{Park2022, Cruz-SaenzdeMiera2022, Cruz-SaenzDeMiera2023}. 
We obtain $\macc = 1.8 \times 10^{-7} - 1.6 \times 10^{-7} \, \msun$yr$^{-1}$, for $d_{min}$; and $\macc = 1.0 - 5.3 \times 10^{-7} \, \msun$yr$^{-1}$, for $d_{BJ}$. 
Both the estimates are similar to the values found in highly accreting CTTSs and in EXors-like objects, e.g. \citet{giannini2022}.

\begin{table}
\centering
\caption{\label{tab:AvLacc}Circumstellar extinction and accretion luminosity estimates for Gaia18cjb using empirical relations from \citet{alc17}.}
\begin{tabular}{lccc}
\hline
\hline
Epoch & A$_V$ & $\log(\lacc/\lsun)$  &  $\log(\lacc/\lsun)$\\
& [mag] & from Pa$\beta$ & from Br$\gamma$\\ 
\hline		
\hline
\\
& & 660 pc \\
\hline
LUCI & $11.0\pm0.3$ & $-0.03 \pm 0.21$ & $-0.04 \pm 0.24$\\
SOFI & $7.9\pm0.3$ & $-0.11 \pm 0.17$ & $-0.11 \pm 0.24$\\
GTC & $12.5\pm0.3$  & $-0.08 \pm 0.17$ & $-0.08 \pm 0.24$\\
\hline
mean & $10.5\pm2.7$ & $-0.07 \pm 0.11$ & $-0.08 \pm 0.14$\\
\hline
\\
& & 1030 pc \\
\hline
LUCI &$11.8\pm0.3$ & $0.39 \pm 0.21$ & $0.39 \pm 0.31$\\
SOFI & $8.7\pm0.3$ & $0.34 \pm 0.21$ & $0.34 \pm 0.32$\\
GTC & $13.3\pm0.3$ & $ 0.33\pm 0.12$ & $ 0.33\pm 0.32$\\
\hline
mean & $11.3\pm0.3$ & $0.35 \pm 0.21$ & $0.35 \pm 0.21$ \\
\hline
\\
& & 7140 pc \\
\hline
LUCI & $15.5\pm0.3$ & $2.68 \pm 0.12$ & $2.67 \pm 0.18$\\
SOFI & $12.4\pm0.3$ & $2.59 \pm 0.12$ & $2.59 \pm 0.18$\\
GTC & $17.0\pm0.3$ & $2.62 \pm 0.12$ & $2.62 \pm 0.18$\\
\hline
mean & $15.0\pm0.3$ & $2.63 \pm 0.07$ & $2.63 \pm 0.10$\\
\hline
\hline
\end{tabular}
 \end{table}

\subsection{On the YSO Nature of Gaia18cjb}
\label{sect:location}
\begin{figure}
    \centering
    \includegraphics[width=\columnwidth]{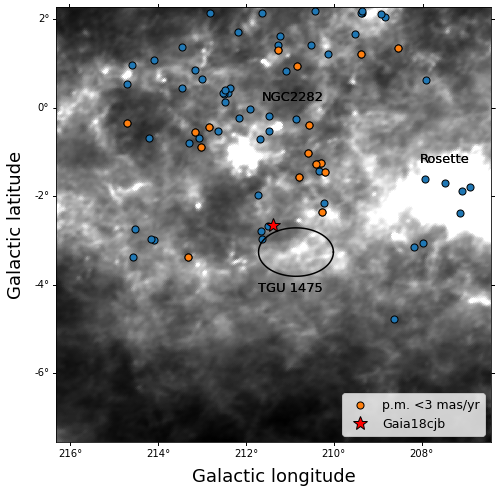}
    \caption{The background of the figure is the Planck 857\,GHz map, 10 degrees in size, centred on Gaia18cjb, the red star symbol in the plot. Blue and orange dots are sources with H$\alpha$-excess from \citet{Fratta2021} whose distances are $0.95\,{\rm pc} < d < 1.05\,{\rm pc}$. The orange dots are sources whose proper motions are within 3\,mas/yr from Gaia18cjb proper motions, see Figure\,\ref{fig:magnitudes}. The regions closes to Gaia18cjb are highlighted.}
    \label{fig:gaia-location}
\end{figure}

\begin{figure*}
    \centering
    \includegraphics[width=\textwidth]{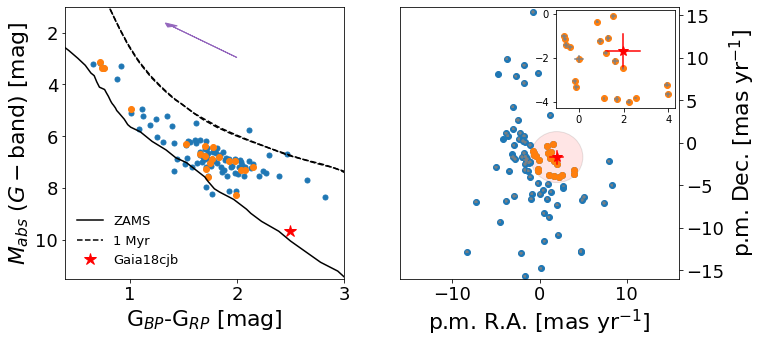}
    \caption{{\it Left:} Color-Absolute Magnitude diagram not corrected for the extinction. The red star symbol corresponds to Gaia18cjb. The black solid and dashed lines show the ZAMS and the 1\,Myr isochrone, respectively \citep{DellOmodarme2013}. Dots are sources which show H$\alpha$-excess from \citet{Fratta2021} catalog whose distance is $ 0.95\,{\rm kpc}\, < d < 1.05$\,kpc. Orange dots are sources with proper motions consistent within $3\sigma$ the propoer motion of Gaia18cjb  (see right panel). The 
error bars smaller than the symbol size are not presented. The purple arrow represents an extinction vector of 1\,mag. However, the $A_V$ vector depends on the color index and on the position in the diagram, see \citet{Prisinzano2022}. {\it Right} Proper motions diagram of the H$\alpha$-excess sample. Symbols are as in the Left pannel.}
    \label{fig:magnitudes}
\end{figure*}

Besides the photometric and spectroscopic features presented in the previous sections,  another important aspect supporting the PMS nature of Gaia18cjb is the enviroment in which the object is located. 
We investigated the region nearby Gaia18cjb looking for its possible association with a star-forming region (SFR). 
Figure\,\ref{fig:gaia-location} shows the environment in the general direction of Gaia18cjb, with the nearby known star-forming regions highlighted. 
Gaia18cjb is located three degrees South-West of the Rosette SFR in the Monoceros constellation. 
Three degrees angular distance corresponds to $\sim50$\,pc at 1\,kp. 
The distance to the Rosette complex is $1489 \pm 32$\,pc \citep{Muzic2022}. 
Thus it is unlikely that Gaia18cjb is connected with the Rosette SFR if its distance is $d \leq d_{BJ}$. Contrary, if Gaia18cjb distance is similar to the Rosette SFR, then it might be part of the star-forming complex.
The VDB\,85 cluster is about $\sim$1.5\,degrees away from Gaia18cjb, and its measured distance is about $\sim$1.69\,kpc \citep{Cantat-Gaudin2020}, so, for the same reasoning applied to the Rosetta Complex, it is unlikely Gaia18cjb belongs to this cluster.
According to the 3D extinction map STILIsM, in correspondence of the TGU\,1475 dark cloud position ($l$, $b = 210.87^\circ$, $-3.28^\circ$), there is one dark cloud at a distance of $\sim 0.9 - 1.0$\,kpc which we may suspect that corresponds to TGU\,1475. 
The ellipse in Fig.\,\ref{fig:gaia-location} shows its nominal size from \citet{Dobashi2005}. 
Although TGU\,1475 and Gaia18cjb could have similar distances under the assumption that $d \leq d_{BJ}$, looking at Fig.\,\ref{fig:gaia-location}, we note that our target is outside the dark cloud. 
Therefore, we may conclude that Gaia18cjb is not a member of any of the known SFRs. 

Recent investigations \citep{Fratta2021, Prisinzano2022}, based on H$\alpha$ surveys and on Gaia DR3, show evidence of populations of YSO candidates distributed in large volumes around the previously known SFRs. 
May Gaia18cjb be part of such population in the general area of Monoceros?
Overplotted in Figure\,\ref{fig:gaia-location} are the sources showing H$\alpha -$excess as selected by \citet{Fratta2021} and whose Gaia distance is between 0.95\,kpc and 1.05\,kpc, similar to the Gaia18cjb distance proposed by \citet{bai21}. 
The lack of H$\alpha$ emitting objects in the southern part is due to the  latitude limit of the INT Photometric H$\alpha$ Survey of the Northern Galactic Plane, IPHAS. 
We also distinguish in this plot objects having proper motions consistent, within $3 \sigma$,to the proper motion of Gaia18cjb (orange dots) and those with inconsistent proper motion (blue dots). 
These two samples are also plotted in the proper motion diagram in Figure\,\ref{fig:magnitudes} (right panel).
Gaia18cjb, shown as a red star symbol in the figure, is not associated with any source in this catalog. 

Figure\,\ref{fig:magnitudes} (left panel) shows a color-magnitude diagram of the sample with H$\alpha$ excess, where the absolute magnitude computed from the Gaia $G-$band as a function of the $G_{BP}-G_{RP}$ color (blue and orange colors are as in Figure\,\ref{fig:gaia-location}). 
The black lines correspond to the ZAMS (solid) and to the 1\,Myr isochrone \citep[dashed, see][]{DellOmodarme2013, Randich2018, Tognelli2018, Tognelli2020}. 
Models have been converted into the Gaia observed plane using the recent EDR3 filters passbands\footnote{GAIA EDR3 passbands are available at the url \url{https://www.cosmos.esa.int/web/gaia/edr3-passbands}.} with the MARCS 2008 \citep{Gustafsson2008} synthetic spectra for T$_\mathrm{eff} \le 8000~$K and \citet{Castelli2003} spectra for higher T$_\mathrm{eff}$ \citep[see e.g.,][]{Prisinzano2022}.
About 97\% (91/94) of the H$\alpha$-excess sample lie above the ZAMS (blue dots), while only three lie below the ZAMS line. 
Yet, many of these objects, in particular those piled-up close to the ZAMS, may be evolved H$\alpha$ emitting stars, like novae and symbiotic stars \citep[see][]{Munari2022}.
Note, however, that symbiotic stars have giant companions and so are located near the giant sequence, and quiescent novae resemble nova-like cataclismatic variables which lie between the white dwarf and main sequences \citep[see, for example][]{Merc2020}. Classical novae in outburst would probably be much brighter than the main sequence.
Note that basically all the objects with proper motion compatible with those of Gaia18cjb (orange dots) lie well above the ZAMS hence, it is likely they are PMS stars. However, wide-band spectroscopy of these objects is necessary to definitely establish their evolutionary status.

The red star symbol lying immediately above the ZAMS corresponds to the Gaia18cjb mean values of absolute magnitude and G$_{\rm BP}- {\rm G}_{\rm RP}$ color provided by the Gaia archive: $G = 19.751 \pm 0.010$\,mag, G$_{\rm BP} = 20.65 \pm 0.77$\,mag, and G$_{\rm RP} = 18.16 \pm 0.77$\,mag. 
These value, reported in Gaia EDR3 are averaged on the period 2014  July $-$ 2017 May, when Gaia18cjb has already started brightening, but was still relatively faint, probably closer to quiescent than to outburst properties.
We stress that the magnitudes of all these sources are not dereddened, implying that after correcting each source for its extinction they will move up and left in the color-magnitude diagram.
For what concerns Gaia18cjb, we estimated an accretion luminosity in burst which is about 27\% of the bolometric luminosity and $A_V \sim 8-12$\,mag. With these values, the contribution of the extinction is larger than the contribution of the accretion, ending up moving Gaia18cjb up in the $M_{\rm abs} \, {\rm vs.} \, {\rm G}_{\rm BP} - {\rm G}_{\rm RP}$ plot.
Therefore, we can speculate that these sources lie above the ZAMS, and are young stars.
The study of the spatial distribution of YSO candidates within 1.5 kpc by \citet{Prisinzano2022},  based on Gaia DR3, highlighted a population of distributed YSO candidates surrounding the  previously known SFRs in the Monoceros region, but Gaia18cjb was not selected by their criterion, probably because too faint to be included in their catalog.

In summary, we selected 20 emission line stars (19 + Gaia18cjb) that: {1}) lie above the ZAMS in the Gaia color-magnitude diagram; {2}) are located at approximately the same  distance ($0.95 \, {\rm kpc} < d < 1.5 \,{\rm kpc}$); {3}) have coherent proper motion. 
We therefore conclude that these likely YSOs, including Gaia18cjb might be part of the distributed population of the Monoceros complex.
This results may indicate that the $d_{BJ}$  could be a good estimate of the real distance of Gaia18cjb. However, as discussed in Sect.\,\ref{App:distance},  the photogeometric distance $d_{BJ}$ may be strongly affected by the prior for the absolute G-band mag, being dominated by the sample of main sequence stars.

Several facts suggest, that Gaia18cjb is a young star: ({\it i}) the high probability of Gaia18cjb to be a YSO: 83\%, computed by \citet{Marton2019}; ({\it ii}) the similarity between Gaia18cjb SED and the SED of a young accreting object (Fig.\,\ref{fig:sed}); ({\it iii}) the spectral index which classifies Gaia18cjb as Class\,I YSO during the pre-burst phase and as a Flat Spectrum YSO during the burst (Sect.\,\ref{sect:SED}); 
({\it iv}) the detection of CTTSs typical features, as accretion and ejection tracers in the NIR spectra (see Sect.\,\ref{sect:emission_lines}); and ({\it v}) the environment in which the object is projected on the sky. 
All these facts together lead us to conclude that Gaia18cjb is an accreting young star.

\subsection{On the Outburst of Gaia18cjb}

The optical light curve (Figure\,\ref{fig:lightcurveVIS}) presents a typical amplitude of a FUor-like object, with timescales compatible with EYOs, longer than typical EXors and smaller than typical FUors. 
On the contrary, the WISE light curves in the MIR show no evidence of such FUor-like strong burst and the variability is smaller than expected for a FUor-like object, while it is compatible with the EXor-like YSOs or even CTTs variability.

If a burst is present, as the optical light curves suggest, it is not clear why it is less pronounced at NIR wavelengths and does not affect the SED at (and possibly beyond) the WISE band 2 (4.6\,$\mu$m), since eruptive objects usually show similar variability at optical and infrared wavelengths \citep[see for example Gaia19fct,][]{Park2022}.
Naively, one could conclude that the large optical and smaller amplitude in the infrared for Gaia18cjb is due to variability in extinction. 
Looking at the color-magnitude diagrams in Figure\,\ref{fig:colordiagrams}, we see that the extinction could only partially explain the variability of Gaia18cjb and, in any case, its low level and high level colors.

For example, if the observed NIR variability ($\Delta W1 \sim 0.6$\,mag, $\Delta m_K \sim 0.7$\,mag, $\Delta m_H \sim 2.2$\,mag, $\Delta m_J \sim 2.9$\,mag) were caused only by extinction, this would correspond to a visible amplitude of $\Delta m_V \sim 10-12$\,mag, that is too large in comparison with the observed visual variability ($\Delta m_r \sim 5$\,mag). 
This suggests that at least some of the observed bluening is intrinsic and not just linked to reddening. 
On the other hand, the accretion burst has to be relatively mild, as the continuum at 4.6\,$\mu$m is barely or not affected at all.
It is thus possible that the observed photometric variability of Gaia18cjb is a combination of enhanced accretion and reduced extinction, as inferred in Gaia19ajj \citep{Hillenbrand2019_19ajj}. The reduced  extinction might possibly rise from winds triggered by the enhanced accretion. 
These would partially clear the dust in the outflow cavities. 
This scenario is compatible with the H$_2$ line seen during the burst (see Sect.\,\ref{sect:emission_lines}).

The burst weakness is also supported by the analysis of the spectra. 
The latter do not show typical FUors-like features like the triangular shape in the $H-$band, the undetected NIR HI and CO lines in absorption \citep{greene_2008}. 
On the contrary, accretion tracers are detected in emission in all the epochs we investigated, confirming this source is an accreting young star and strengthening the EXor nature of Gaia18cjb burst. 

\begin{figure}
    \centering
    \includegraphics[width=\columnwidth]{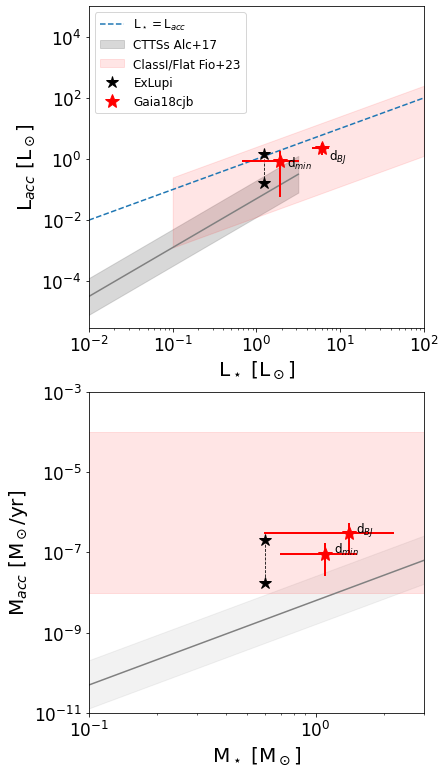}
    \caption{{\it Top:} $\lacc - \lstar$ diagram. The red and black star symbols correspond to Gaia18cjb and EX\,Lup, respectively. The blue dashed line represents the $\lacc = \lstar$ locus. The grey line is the best fit for the CTTSs of Lupus from \citet{alc17} and the grey region corresponds to the standard deviation of the ﬁt.
    {\it Bottom:} $\macc - \mstar$ diagram. Star symbols are as in the top panel. The grey line is the best fit for the CTTSs of Lupus from \citet{alc17} and the grey region corresponds to the standard deviation of the ﬁt. The pink area corresponds to $\macc$ ranges for embedded Class\,I+Flat Spectrum YSOs from \citet{Fiorellino2023}. 
    }
    \label{fig:Lacc-Lstar}
\end{figure}

In Sect.\,\ref{sect:stellar_luminosity}, using empirical relations, we find that $\lacc$ ranges from 0.85\,$\lsun$ to 2.24\,$\lsun$, and $\macc$ ranges from $2.8 \times 10^{-8}\, \msun/$yr to $1.0 \times 10^{-7} \, \msun/$yr if $d_{min} \leq d \leq d_{BJ}$.

Calculating $\lacc$ using empirical relationships from \citet{alc17} is an indirect method that has shown success with CTTSs but some evidence has recently emerged that it may not be always applicable to eruptive sources, even if we expect magnetospheric accretion, such as for EXors objects. 
For instance, recent analysis of EX\,Lupi revealed that several emission lines indicated lower values of $\lacc$ compared to the slab model, the direct way to determine $\lacc$ in the framework of magnetospheric accretion \citep{Cruz-SaenzDeMiera2023}. 
They suggest that this discrepancy may arise from the challenge of identifying the continuum in a highly accreting objects. 
To investigate whether this holds true for Gaia18cjb, we would require wide-band spectroscopic data covering the Balmer continuum and Balmer jump for application of a slab model fit.
Since we lack such spectroscopic data, and conversely, we can only calculate $\lacc$ from the $\pab$ and $\brg$ lines in the NIR, the estimate of $\lacc$ obtained from empirical relationships using these lines is currently the best estimate of $\lacc$. 
However, we emphasize that due to the likely overestimated value of Gaia18cjb's stellar luminosity and the uncertainty on the distance, there is a high probability that the accretion luminosity exceeds $\lacc = 2.24 \, \lsun$.  
This needs to be tested through the direct measurement of $\lacc$ by studying the Balmer continuum excess emission and Balmer jump, using wide-band spectra. 

Figure\,\ref{fig:Lacc-Lstar} shows the $\lacc \,{\rm vs.}\, \lstar$ (top panel) and $\macc \,{\rm vs.}\, \mstar$ (bottom panel) of Gaia18cjb at $d_{min} = 660$\,pc and $d_{BJ} = 1.03$\,kpc (red star symbols), compared to EX\,Lup  \citep[black stars, quiescent value from \citet{Cruz-SaenzDeMiera2023}, and burst value from][]{fis22pp7} and typical distributions for CTTSs in Lupus \citep[grey,][]{alc17}, and Class\,I/Flat\,Spectrum protostars \citep[pink,][]{Fiorellino2023}. 

The top panel of Figure\,\ref{fig:Lacc-Lstar} shows that Gaia18cjb is located in the Class\,I/Flat\,Spectrum YSOs parameter space, right below the $\lacc = \lstar$ threshold (dashed blue line). 
Therefore, during the epochs we investigated with spectroscopy, i.e. during the burst, the ratio between the stellar and the accretion luminosity was more similar to a YSO more evolved  than a protostar  if $d_{min} \leq d \leq d_{BJ}$, and its accretion rate and stellar parameters are compatible with a Class\,I/Flat Spectrum object, suggesting Gaia18cjb evolutionary stage is in between the protostellar and the pre-main sequence phases. This in agreement with the Flat Spectrum classification suggested by the spectral index (Sect.\,\ref{sect:SED}).
Note, however, that the luminosity of Gaia18cjb is still compatible with that of typical CTTS if the distance $d_{min} = 660$\,pc is adopted.

Regarding the $\macc$\,vs.\,$\mstar$ plot, the bottom panel of Figure\,\ref{fig:Lacc-Lstar} shows that Gaia18cjb accretes more than typical CTTSs of the same stellar mass. 
Gaia18cjb location in this plot is in agreement with bright Class\,I and Flat Spectrum sources, and the mass accretion rate is similar to what derived for EX\,Lup during the burst. 
Indeed, while the optical light curve resemble a FUor, the spectroscopic features and the accretion rate suggest an EXor-like nature of Gaia18cjb.
Interestingly, there are at least two young stars, V350\,Cep and V1647\,Ori, displaying photometric characteristics of FUors in the visible bands and spectral characteristics of EXors \citep{Herbig2008,Ibryamov2014,Aspin2009}. 
In both cases, the conclusion was that the sources were probably intermediate objects between FUors and EXors. 

The similarities between V350\,Cep, V1647\,Ori, and Gaia18cjb, might suggest that also Gaia18cjb is an intermediate object between FUors and EXors.
However, there are a few aspects to take into account.
First, V350\,Cep and V1647\,Ori also show P\,Cyg H$\alpha$ profile, and random fluctuations in brightness with amplitudes of a few tenths of a magnitude in timescales of several days that we do not see in Gaia18cjb. 
Second, the current discussion about EYSs classification highlights that the classification in very tight containers (e.g. FUor, EXor, V1647 Or) is not sufficient. 
Gaia18cjb is another star suggesting instead a spectrum of classifications in the EYSs.
Last, our results about accretion rates and stellar parameters should be considered as rough estimates, because our analysis is based only on NIR spectroscopy and because of the large uncertainty on the distance. 
To provide accurate parameters and hence, a robust characterization of Gaia18cjb, further observations and analysis are needed: 
\begin{itemize}

\item Continuing monitoring optical and infrared light curves is crucial to follow the evolution of the Gaia18cjb burst. Also, $JHK$ NIR infrared monitoring is needed to study the light curve variability in between the FUor-like visible light curve and the EXor-like infrared light curves. 

\item By having a flux calibrated intermediate/high-resolution spectrum from the UVB to the NIR (as ESO/VLT XShooter can provide), we will be able to study the Li\,I ($\lambda$6708\,\AA) absorption line to reinforce the YSOs nature of Gaia18cjb, to perform a direct spectral typing and determine in a more accurate way the stellar parameters, and hence, measure the accretion luminosity from the Balmer continuum excess emission. 
As a consequence, we will be able to provide robust estimates of stellar and accretion parameters, and of the mass accretion rate. 

\item Millimeter interferometry will provide estimates for disk mass and radius which are needed to prove the instability of the disk, believed to trigger the eruptive accretion.

\item Accurate distance estimate, by using a pre-main-sequence dominated prior in the \citet{bai21} prescription, or a more accurate parallax estimate from GDR4.
\end{itemize}


\section{Conclusions}
\label{sect:conclusions}

The goal of this discovery paper was to determine the nature of the Gaia alerted object Gaia18cjb.

For this purpose, we collected all the available photometry in the archives and monitored Gaia18cjb in optical bands. 
NIR spectroscopy was also performed in three epochs between 2020 and 2022 and we studied the object's variability, stellar and accretion parameters, and inspected the sky region nearby Gaia18cjb.

The large uncertainty on the Gaia parallax and the main-sequence dominated prior of the \citet{bai21} method, prevent us from adopting a precise distance for this source. 
Because of this, we cannot derive precise values but only ranges of stellar and accretion parameters based on the comparison between analytical results and spectral features.
Comparing the Gaia stellar parameters and accretion rates with those of CTTSs, protostars, FUors, and other similar eruptive objects as EX\,Lup, V350\,Cep, V1647\,Ori, and Gaia19ajj, our results suggest Gaia18cjb is a young star experiencing eruptive accretion and that it can be classified as in between FUors and EXors: the length of the burst points to a FUor-like nature, whereas the strength of the burst and detected spectral features point to an EXor.
However, further optical and NIR monitoring campaigns, at least one intermediate/high-resolution spectrum from UVB to NIR, and  millimeter interferometry, are needed to determine accurate stellar and accretion parameters and, hence, provide robust conclusions on the Gaia18cjb variability classification.

\begin{acknowledgements}
We thank Dr. C. Bailer Jones for discussions on the parallax of the object and Dr. L. Prisinzano for providing the isochrones in the Gaia photometric system. 
We acknowledge Dr. Andr\'as Ordasi for his contribution to this paper. 
This work has been supported by the
project PRIN‐INAF 2019 “Spectroscopically Tracing the Disk
Dispersal Evolution (STRADE)” and by the INAF Large Grant 2022 “YSOs Outflow, Disks and Accretion (YODA)”.

This project has received funding from the European Research Council (ERC) under the European Union's Horizon 2020 research and innovation programme under grant agreement No 716155 (SACCRED). 

We acknowledge ESA Gaia, DPAC and the Photometric Science Alerts Team (\url{http://gsaweb.ast.cam.ac.uk/alerts})

This work is (partly) based on data obtained with the instrument EMIR,
built by a Consortium led by the Instituto de Astrofísica de Canarias.
EMIR was funded by GRANTECAN and the National Plan of Astronomy and
Astrophysics of the Spanish Government. 
This work is (partly) based on data obtained at the Mount Suhora Observatory, Krakow Pedagogical University, Poland.
This work has been supported by the Hungarian National Research, Development and Innovation Office grants OTKA K131508, OTKA K138962, the \'Elvonal KKP-143986 and KKP-137523 'SeismoLab' grants of the Hungarian Research, Development and Innovation Office (NKFIH).
We acknowledge support from the ESA PRODEX contract nr. 4000132054.
KV, LK, ZN, and GM are supported by the Bolyai J\'anos Research Scholarship of the Hungarian Academy of Sciences, KV is supported by the Bolyai+ grant \'UNKP-22-5-ELTE-1093. Authors acknowledge the financial support of the Austrian-Hungarian Action Foundation (112\"ou1). LK is supported by the Hungarian National Research, Development and Innovation Office grant PD-134784.
G.M. acknowledges support from the European Union’s Horizon 2020 research and innovation programme under grant agreement No. 101004141. 
This work is supported by the Polish MNiSW grant DIR/WK/2018/12 and the European Union's Horizon 2020 research and innovation program under grant agreement No.\,101004719 (OPTICON-RadioNet Pilot). 
Zs.M.Sz. acknowledges funding from a St Leonards scholarship from the University of St Andrews.
For the purpose of open access, the author has applied a Creative Commons
Attribution (CC BY) licence to any Author Accepted Manuscript version arising. 
BZs is supported by the ÚNKP-22-2 New National Excellence Program of the Ministry for Culture and Innovation from the source of the National Research, Development and Innovation Fund. 
The Liverpool Telescope is operated on the island of La Palma by Liverpool John Moores University in the Spanish Observatorio del Roque de los Muchachos of the Instituto de Astrofisica de Canarias with financial support from the UK Science and Technology Facilities Council. We acknowledge the Hungarian National Research, Development and Innovation Office grant OTKA FK 146023. FCSM received financial support from the European Research Council (ERC) under the European Union’s Horizon 2020 research and innovation programme (ERC Starting Grant “Chemtrip", grant agreement No 949278).

\end{acknowledgements}

\bibliographystyle{aa} 
\bibliography{biblio.bib}

\begin{thebibliography}{84}
\expandafter\ifx\csname natexlab\endcsname\relax\def\natexlab#1{#1}\fi

\bibitem[{{Alcal{\'a}} {et~al.}(2017){Alcal{\'a}}, {Manara}, {Natta}, {Frasca},
  {Testi}, {Nisini}, {Stelzer}, {Williams}, {Antoniucci}, \& {Biazzo}}]{alc17}
{Alcal{\'a}}, J.~M., {Manara}, C.~F., {Natta}, A., {et~al.} 2017, \aap, 600,
  A20

\bibitem[{{Andr{é}}(1995)}]{and95}
{Andr{é}}, P. 1995, \apss, 224, 29

\bibitem[{{Antoniucci} {et~al.}(2008){Antoniucci}, {Nisini}, {Giannini}, \&
  {Lorenzetti}}]{ant08}
{Antoniucci}, S., {Nisini}, B., {Giannini}, T., \& {Lorenzetti}, D. 2008, \aap,
  479, 503

\bibitem[{{Arnaboldi} {et~al.}(2016){Arnaboldi}, {Delmotte}, {Geier}, {Hilker},
  {Hussain}, {Mascetti}, {Micol}, {Petr-Gotzens}, {Rejkuba}, \&
  {Retzlaff}}]{Arnaboldi2016}
{Arnaboldi}, M., {Delmotte}, N., {Geier}, S., {et~al.} 2016, in Astrophysics
  and Space Science Proceedings, Vol.~42, The Universe of Digital Sky Surveys,
  ed. N.~R. {Napolitano}, G.~{Longo}, M.~{Marconi}, M.~{Paolillo}, \&
  E.~{Iodice}, 25

\bibitem[{{Aspin} \& {Reipurth}(2009)}]{Aspin2009}
{Aspin}, C. \& {Reipurth}, B. 2009, \aj, 138, 1137

\bibitem[{{Audard} {et~al.}(2014){Audard}, {{\'A}brah{\'a}m}, {Dunham},
  {Green}, {Grosso}, {Hamaguchi}, {Kastner}, {K{\'o}sp{\'a}l}, {Lodato},
  {Romanova}, {Skinner}, {Vorobyov}, \& {Zhu}}]{audard2014}
{Audard}, M., {{\'A}brah{\'a}m}, P., {Dunham}, M.~M., {et~al.} 2014, in
  Protostars and Planets VI, ed. H.~{Beuther}, R.~S. {Klessen}, C.~P.
  {Dullemond}, \& T.~{Henning}, 387--410

\bibitem[{{Bailer-Jones} {et~al.}(2021){Bailer-Jones}, {Rybizki}, {Fouesneau},
  {Demleitner}, \& {Andrae}}]{bai21}
{Bailer-Jones}, C.~A.~L., {Rybizki}, J., {Fouesneau}, M., {Demleitner}, M., \&
  {Andrae}, R. 2021, \aj, 161, 147

\bibitem[{{Cantat-Gaudin} \& {Anders}(2020)}]{Cantat-Gaudin2020}
{Cantat-Gaudin}, T. \& {Anders}, F. 2020, \aap, 633, A99

\bibitem[{{Capitanio} {et~al.}(2017){Capitanio}, {Lallement}, {Vergely},
  {Elyajouri}, \& {Monreal-Ibero}}]{Capitanio2017}
{Capitanio}, L., {Lallement}, R., {Vergely}, J.~L., {Elyajouri}, M., \&
  {Monreal-Ibero}, A. 2017, \aap, 606, A65

\bibitem[{{Cardelli} {et~al.}(1989){Cardelli}, {Clayton}, \& {Mathis}}]{car89}
{Cardelli}, J.~A., {Clayton}, G.~C., \& {Mathis}, J.~S. 1989, \apj, 345, 245

\bibitem[{{Carpenter} {et~al.}(2001){Carpenter}, {Hillenbrand}, \&
  {Skrutskie}}]{car01}
{Carpenter}, J.~M., {Hillenbrand}, L.~A., \& {Skrutskie}, M.~F. 2001, \aj, 121,
  3160

\bibitem[{{Castelli} \& {Kurucz}(2003)}]{Castelli2003}
{Castelli}, F. \& {Kurucz}, R.~L. 2003, in Modelling of Stellar Atmospheres,
  ed. N.~{Piskunov}, W.~W. {Weiss}, \& D.~F. {Gray}, Vol. 210, A20

\bibitem[{{Cody} {et~al.}(2014){Cody}, {Stauffer}, {Baglin}, {Micela},
  {Rebull}, {Flaccomio}, {Morales-Calder{\'o}n}, {Aigrain}, {Bouvier},
  {Hillenbrand}, {Gutermuth}, {Song}, {Turner}, {Alencar}, {Zwintz},
  {Plavchan}, {Carpenter}, {Findeisen}, {Carey}, {Terebey}, {Hartmann},
  {Calvet}, {Teixeira}, {Vrba}, {Wolk}, {Covey}, {Poppenhaeger}, {G{\"u}nther},
  {Forbrich}, {Whitney}, {Affer}, {Herbst}, {Hora}, {Barrado}, {Holtzman},
  {Marchis}, {Wood}, {Medeiros Guimar{\~a}es}, {Lillo Box}, {Gillen},
  {McQuillan}, {Espaillat}, {Allen}, {D'Alessio}, \& {Favata}}]{cody2014}
{Cody}, A.~M., {Stauffer}, J., {Baglin}, A., {et~al.} 2014, \aj, 147, 82

\bibitem[{{Connelley} \& {Reipurth}(2018)}]{connelley-reipurth2018}
{Connelley}, M.~S. \& {Reipurth}, B. 2018, \apj, 861, 145

\bibitem[{{Cruz-S{\'a}enz de Miera} {et~al.}(2023){Cruz-S{\'a}enz de Miera},
  {K{\'o}sp{\'a}l}, {{\'A}brah{\'a}m}, {Claes}, {Manara}, {Wendeborn},
  {Fiorellino}, {Giannini}, {Nisini}, {Sicilia-Aguilar}, {Campbell-White},
  {Alcal{\'a}}, {Banzatti}, {Szab{\'o}}, {Lykou}, {Antoniucci}, {Varga},
  {Siwak}, {Park}, {Nagy}, \& {Kun}}]{Cruz-SaenzDeMiera2023}
{Cruz-S{\'a}enz de Miera}, F., {K{\'o}sp{\'a}l}, {\'A}., {{\'A}brah{\'a}m}, P.,
  {et~al.} 2023, arXiv e-prints, arXiv:2308.02849

\bibitem[{{Cruz-S{\'a}enz de Miera} {et~al.}(2022){Cruz-S{\'a}enz de Miera},
  {K{\'o}sp{\'a}l}, {{\'A}brah{\'a}m}, {Park}, {Nagy}, {Siwak}, {Kun},
  {Fiorellino}, {Szab{\'o}}, {Antoniucci}, {Giannini}, {Nisini}, {Szabados},
  {Kriskovics}, {Ordasi}, {Szak{\'a}ts}, {Vida}, {Vink{\'o}}, {Zieli{\'n}ski},
  {Wyrzykowski}, {Garc{\'\i}a-{\'A}lvarez}, {Dr{\'o}{\.z}d{\.z}}, {Og{\l}oza},
  \& {Sonbas}}]{Cruz-SaenzdeMiera2022}
{Cruz-S{\'a}enz de Miera}, F., {K{\'o}sp{\'a}l}, {\'A}., {{\'A}brah{\'a}m}, P.,
  {et~al.} 2022, \apj, 927, 125

\bibitem[{{Cutri} {et~al.}(2003){Cutri}, {Skrutskie}, {van Dyk}, {Beichman},
  {Carpenter}, {Chester}, {Cambresy}, {Evans}, {Fowler}, {Gizis}, {Howard},
  {Huchra}, {Jarrett}, {Kopan}, {Kirkpatrick}, {Light}, {Marsh}, {McCallon},
  {Schneider}, {Stiening}, {Sykes}, {Weinberg}, {Wheaton}, {Wheelock}, \&
  {Zacarias}}]{cutri2003}
{Cutri}, R.~M., {Skrutskie}, M.~F., {van Dyk}, S., {et~al.} 2003, VizieR Online
  Data Catalog, II/246

\bibitem[{{Dell'Omodarme} \& {Valle}(2013)}]{DellOmodarme2013}
{Dell'Omodarme}, M. \& {Valle}, G. 2013, arXiv e-prints, arXiv:1301.3695

\bibitem[{{Dobashi} {et~al.}(2005){Dobashi}, {Uehara}, {Kandori}, {Sakurai},
  {Kaiden}, {Umemoto}, \& {Sato}}]{Dobashi2005}
{Dobashi}, K., {Uehara}, H., {Kandori}, R., {et~al.} 2005, VizieR Online Data
  Catalog, VII/244A

\bibitem[{{Evans} {et~al.}(2009){Evans}, {Dunham}, {Jorgensen}, {Enoch},
  {Merin}, {van Dishoeck}, {Alcala}, {Myers}, {Stapelfeldt}, \&
  {Huard}}]{eva09}
{Evans}, N.~J., {Dunham}, M.~M., {Jorgensen}, J.~K., {et~al.} 2009, VizieR
  Online Data Catalog, J/ApJS/181/321

\bibitem[{{Fiorellino} {et~al.}(2021){Fiorellino}, {Manara}, {Nisini},
  {Ramsay}, {Antoniucci}, {Giannini}, {Biazzo}, {Alcal{\`a}}, \&
  {Fedele}}]{fio21}
{Fiorellino}, E., {Manara}, C.~F., {Nisini}, B., {et~al.} 2021, \aap, 650, A43

\bibitem[{{Fiorellino} {et~al.}(2023){Fiorellino}, {Tychoniec}, {Cruz-S{\'a}enz
  de Miera}, {Antoniucci}, {K{\'o}sp{\'a}l}, {Manara}, {Nisini}, \&
  {Rosotti}}]{Fiorellino2023}
{Fiorellino}, E., {Tychoniec}, {\L}., {Cruz-S{\'a}enz de Miera}, F., {et~al.}
  2023, \apj, 944, 135

\bibitem[{{Fiorellino} {et~al.}(2022){Fiorellino}, {Zsidi}, {K{\'o}sp{\'a}l},
  {{\'A}brah{\'a}m}, {B{\'o}di}, {Hussain}, {Manara}, \&
  {P{\'a}l}}]{fiorellino2022wx}
{Fiorellino}, E., {Zsidi}, G., {K{\'o}sp{\'a}l}, {\'A}., {et~al.} 2022, \apj,
  938, 93

\bibitem[{{Fischer} {et~al.}(2022){Fischer}, {Hillenbrand}, {Herczeg},
  {Johnstone}, {K{\'o}sp{\'a}l}, \& {Dunham}}]{fis22pp7}
{Fischer}, W.~J., {Hillenbrand}, L.~A., {Herczeg}, G.~J., {et~al.} 2022, arXiv
  e-prints, arXiv:2203.11257

\bibitem[{{Fratta} {et~al.}(2021){Fratta}, {Scaringi}, {Drew}, {Monguio},
  {Knigge}, {Maccarone}, {Court}, {Ilkiewicz}, {Pala}, {Gandhi}, \&
  {Gaensicke}}]{Fratta2021}
{Fratta}, M., {Scaringi}, S., {Drew}, J.~E., {et~al.} 2021, VizieR Online Data
  Catalog, J/MNRAS/505/1135

\bibitem[{{Gargiulo} {et~al.}(2022){Gargiulo}, {Fumana}, {Bisogni},
  {Franzetti}, {Cassar{\`a}}, {Garilli}, {Scodeggio}, \&
  {Vietri}}]{Gargiulo2022}
{Gargiulo}, A., {Fumana}, M., {Bisogni}, S., {et~al.} 2022, \mnras, 514, 2902

\bibitem[{{Garz{\'o}n} {et~al.}(2022){Garz{\'o}n}, {Balcells}, {Gallego},
  {Gry}, {Guzm{\'a}n}, {Hammersley}, {Herrero}, {Mu{\~n}oz-Tu{\~n}{\'o}n},
  {Pell{\'o}}, {Prieto}, {Bourrec}, {Cabello}, {Cardiel},
  {Gonz{\'a}lez-Fern{\'a}ndez}, {Laporte}, {Milliard}, {Pascual}, {Patrick},
  {Patr{\'o}n}, {Ram{\'\i}rez-Alegr{\'\i}a}, \& {Streblyanska}}]{Garzon2022}
{Garz{\'o}n}, F., {Balcells}, M., {Gallego}, J., {et~al.} 2022, \aap, 667, A107

\bibitem[{{Ghosh} {et~al.}(2022){Ghosh}, {Sharma}, {Ninan}, {Ojha}, {Bhatt},
  {Kanodia}, {Mahadevan}, {Stefansson}, {Yadav}, {Gour}, {Pandey}, {Sinha},
  {Panwar}, {Wisniewski}, {Ca{\~n}as}, {Lin}, {Roy}, {Hearty}, {Ramsey},
  {Robertson}, \& {Schwab}}]{Ghosh2022}
{Ghosh}, A., {Sharma}, S., {Ninan}, J.~P., {et~al.} 2022, \apj, 926, 68

\bibitem[{{Giannini} {et~al.}(2022){Giannini}, {Giunta}, {Gangi}, {Carini},
  {Lorenzetti}, {Antoniucci}, {Caratti o Garatti}, {Cassar{\'a}}, {Nisini},
  {Rossi}, {Testa}, \& {Vitali}}]{giannini2022}
{Giannini}, T., {Giunta}, A., {Gangi}, M., {et~al.} 2022, \apj, 929, 129

\bibitem[{{Green} {et~al.}(2019){Green}, {Schlafly}, {Zucker}, {Speagle}, \&
  {Finkbeiner}}]{green_2019}
{Green}, G.~M., {Schlafly}, E., {Zucker}, C., {Speagle}, J.~S., \&
  {Finkbeiner}, D. 2019, \apj, 887, 93

\bibitem[{{Greene} {et~al.}(2008){Greene}, {Aspin}, \&
  {Reipurth}}]{greene_2008}
{Greene}, T.~P., {Aspin}, C., \& {Reipurth}, B. 2008, \aj, 135, 1421

\bibitem[{{Greene} {et~al.}(1994){Greene}, {Wilking}, {Andre}, {Young}, \&
  {Lada}}]{Greene1994}
{Greene}, T.~P., {Wilking}, B.~A., {Andre}, P., {Young}, E.~T., \& {Lada},
  C.~J. 1994, \apj, 434, 614

\bibitem[{{Gustafsson} {et~al.}(2008){Gustafsson}, {Edvardsson}, {Eriksson},
  {J{\o}rgensen}, {Nordlund}, \& {Plez}}]{Gustafsson2008}
{Gustafsson}, B., {Edvardsson}, B., {Eriksson}, K., {et~al.} 2008, \aap, 486,
  951

\bibitem[{{Hartmann} {et~al.}(1998){Hartmann}, {Calvet}, {Gullbring}, \&
  {D'Alessio}}]{har98}
{Hartmann}, L., {Calvet}, N., {Gullbring}, E., \& {D'Alessio}, P. 1998, \apj,
  495, 385

\bibitem[{{Hartmann} {et~al.}(2016){Hartmann}, {Herczeg}, \& {Calvet}}]{har16}
{Hartmann}, L., {Herczeg}, G., \& {Calvet}, N. 2016, \araa, 54, 135

\bibitem[{{Hartmann} \& {Kenyon}(1996)}]{har96}
{Hartmann}, L. \& {Kenyon}, S.~J. 1996, \araa, 34, 207

\bibitem[{{Henden} {et~al.}(2016){Henden}, {Templeton}, {Terrell}, {Smith},
  {Levine}, \& {Welch}}]{henden2016}
{Henden}, A.~A., {Templeton}, M., {Terrell}, D., {et~al.} 2016, VizieR Online
  Data Catalog, II/336

\bibitem[{{Herbig}(1977)}]{Herbig77}
{Herbig}, G.~H. 1977, \apj, 217, 693

\bibitem[{{Herbig}(1989)}]{Herbig89}
{Herbig}, G.~H. 1989, in European Southern Observatory Conference and Workshop
  Proceedings, Vol.~33, European Southern Observatory Conference and Workshop
  Proceedings, 233--246

\bibitem[{{Herbig}(2008)}]{Herbig2008}
{Herbig}, G.~H. 2008, \aj, 135, 637

\bibitem[{{Hillenbrand} {et~al.}(2018){Hillenbrand}, {Contreras Pe{\~n}a},
  {Morrell}, {Naylor}, {Kuhn}, {cutri}, {Rebull}, {Hodgkin}, {Froebrich}, \&
  {Mainzer}}]{hillenbrand2018}
{Hillenbrand}, L.~A., {Contreras Pe{\~n}a}, C., {Morrell}, S., {et~al.} 2018,
  \apj, 869, 146

\bibitem[{{Hillenbrand} {et~al.}(2019{\natexlab{a}}){Hillenbrand}, {Reipurth},
  {Connelley}, {cutri}, \& {Isaacson}}]{hillenbrand2019}
{Hillenbrand}, L.~A., {Reipurth}, B., {Connelley}, M., {cutri}, R.~M., \&
  {Isaacson}, H. 2019{\natexlab{a}}, \aj, 158, 240

\bibitem[{{Hillenbrand} {et~al.}(2019{\natexlab{b}}){Hillenbrand}, {Reipurth},
  {Connelley}, {Cutri}, \& {Isaacson}}]{Hillenbrand2019_19ajj}
{Hillenbrand}, L.~A., {Reipurth}, B., {Connelley}, M., {Cutri}, R.~M., \&
  {Isaacson}, H. 2019{\natexlab{b}}, \aj, 158, 240

\bibitem[{{Hodapp} {et~al.}(2020){Hodapp}, {Denneau}, {Tucker}, {Shappee},
  {Huber}, {Payne}, {Do}, {Lin}, {Connelley}, {Varricatt}, {Tonry}, {Chambers},
  \& {Magnier}}]{hodapp2020}
{Hodapp}, K.~W., {Denneau}, L., {Tucker}, M., {et~al.} 2020, \aj, 160, 164

\bibitem[{{Hodapp} {et~al.}(2019){Hodapp}, {Reipurth}, {Pettersson}, {Tonry},
  {Denneau}, {Vallely}, {Shappee}, {Armstrong}, {Connelley}, {Kochanek},
  {Fausnaugh}, {Chini}, {Haas}, \& {Sobrino Figaredo}}]{Hodapp2019}
{Hodapp}, K.~W., {Reipurth}, B., {Pettersson}, B., {et~al.} 2019, \aj, 158, 241

\bibitem[{{Hodgkin} {et~al.}(2021){Hodgkin}, {Harrison}, {Breedt}, {Wevers},
  {Rixon}, {Delgado}, {Yoldas}, {Kostrzewa-Rutkowska}, {Wyrzykowski}, {van
  Leeuwen}, {Blagorodnova}, {Campbell}, {Eappachen}, {Fraser}, {Ihanec},
  {Koposov}, {Kruszy{\'n}ska}, {Marton}, {Rybicki}, {Brown}, {Burgess},
  {Busso}, {Cowell}, {De Angeli}, {Diener}, {Evans}, {Gilmore}, {Holland},
  {Jonker}, {van Leeuwen}, {Mignard}, {Osborne}, {Portell}, {Prusti},
  {Richards}, {Riello}, {Seabroke}, {Walton}, {{\'A}brah{\'a}m}, {Altavilla},
  {Baker}, {Bastian}, {O'Brien}, {de Bruijne}, {Butterley}, {Carrasco},
  {Casta{\~n}eda}, {Clark}, {Clementini}, {Copperwheat}, {Cropper},
  {Damljanovic}, {Davidson}, {Davis}, {Dennefeld}, {Dhillon}, {Dolding},
  {Dominik}, {Esquej}, {Eyer}, {Fabricius}, {Fridman}, {Froebrich}, {Garralda},
  {Gomboc}, {Gonz{\'a}lez-Vidal}, {Guerra}, {Hambly}, {Hardy}, {Holl},
  {Hourihane}, {Japelj}, {Kann}, {Kiss}, {Knigge}, {Kolb}, {Komossa},
  {K{\'o}sp{\'a}l}, {Kov{\'a}cs}, {Kun}, {Leto}, {Lewis}, {Littlefair},
  {Mahabal}, {Mundell}, {Nagy}, {Padeletti}, {Palaversa}, {Pigulski},
  {Pretorius}, {van Reeven}, {Ribeiro}, {Roelens}, {Rowell}, {Schartel},
  {Scholz}, {Schwope}, {Sip{\H{o}}cz}, {Smartt}, {Smith}, {Serraller},
  {Steeghs}, {Sullivan}, {Szabados}, {Szegedi-Elek}, {Tisserand}, {Tomasella},
  {van Velzen}, {Whitelock}, {Wilson}, \& {Young}}]{Hodgkin2021}
{Hodgkin}, S.~T., {Harrison}, D.~L., {Breedt}, E., {et~al.} 2021, \aap, 652,
  A76

\bibitem[{{Ibryamov} {et~al.}(2014){Ibryamov}, {Semkov}, \&
  {Peneva}}]{Ibryamov2014}
{Ibryamov}, S., {Semkov}, E., \& {Peneva}, S. 2014, Research in Astronomy and
  Astrophysics, 14, 1264

\bibitem[{{Kenyon} {et~al.}(1990){Kenyon}, {Hartmann}, {Strom}, \&
  {Strom}}]{ken90}
{Kenyon}, S.~J., {Hartmann}, L.~W., {Strom}, K.~M., \& {Strom}, S.~E. 1990,
  \aj, 99, 869

\bibitem[{{K{\'o}sp{\'a}l} {et~al.}(2011){K{\'o}sp{\'a}l}, {{\'A}brah{\'a}m},
  {Acosta-Pulido}, {Ar{\'e}valo Morales}, {Carnerero}, {Elek}, {Kelemen},
  {Kun}, {P{\'a}l}, {Szak{\'a}ts}, \& {Vida}}]{kospal2011}
{K{\'o}sp{\'a}l}, {\'A}., {{\'A}brah{\'a}m}, P., {Acosta-Pulido}, J.~A.,
  {et~al.} 2011, \aap, 527, A133

\bibitem[{{Lada}(1987)}]{lad87}
{Lada}, C.~J. 1987, in IAU Symposium, Vol. 115, Star Forming Regions, ed.
  M.~{Peimbert} \& J.~{Jugaku}, 1

\bibitem[{{Lada} \& {Wilking}(1984)}]{lad84}
{Lada}, C.~J. \& {Wilking}, B.~A. 1984, \apj, 287, 610

\bibitem[{{Lallement} {et~al.}(2018){Lallement}, {Capitanio}, {Ruiz-Dern},
  {Danielski}, {Babusiaux}, {Vergely}, {Elyajouri}, {Arenou}, \&
  {Leclerc}}]{Lallement2018}
{Lallement}, R., {Capitanio}, L., {Ruiz-Dern}, L., {et~al.} 2018, \aap, 616,
  A132

\bibitem[{{Lorenzetti} {et~al.}(2012){Lorenzetti}, {Antoniucci}, {Giannini},
  {Li Causi}, {Ventura}, {Arkharov}, {Kopatskaya}, {Larionov}, {Di Paola}, \&
  {Nisini}}]{lorenzetti2012}
{Lorenzetti}, D., {Antoniucci}, S., {Giannini}, T., {et~al.} 2012, \apj, 749,
  188

\bibitem[{{Lorenzetti} {et~al.}(2011){Lorenzetti}, {Giannini}, {Larionov},
  {Arkharov}, {Antoniucci}, {Di Paola}, {Konstantinova}, {Kopatskaya}, {Li
  Causi}, \& {Nisini}}]{lorenzetti2011}
{Lorenzetti}, D., {Giannini}, T., {Larionov}, V.~M., {et~al.} 2011, \apj, 732,
  69

\bibitem[{{Lorenzetti} {et~al.}(2007){Lorenzetti}, {Giannini}, {Larionov},
  {Kopatskaya}, {Arkharov}, {De Luca}, \& {Di Paola}}]{lorenzetti2007}
{Lorenzetti}, D., {Giannini}, T., {Larionov}, V.~M., {et~al.} 2007, \apj, 665,
  1182

\bibitem[{{Lucas} {et~al.}(2008){Lucas}, {Hoare}, {Longmore}, {Schr{\"o}der},
  {Davis}, {Adamson}, {Bandyopadhyay}, {de Grijs}, {Smith}, {Gosling},
  {Mitchison}, {G{\'a}sp{\'a}r}, {Coe}, {Tamura}, {Parker}, {Irwin}, {Hambly},
  {Bryant}, {Collins}, {Cross}, {Evans}, {Gonzalez-Solares}, {Hodgkin},
  {Lewis}, {Read}, {Riello}, {Sutorius}, {Lawrence}, {Drew}, {Dye}, \&
  {Thompson}}]{lucas2008}
{Lucas}, P.~W., {Hoare}, M.~G., {Longmore}, A., {et~al.} 2008, \mnras, 391, 136

\bibitem[{{Mainzer} {et~al.}(2014){Mainzer}, {Bauer}, {cutri}, {Grav},
  {Masiero}, {Beck}, {Clarkson}, {Conrow}, {Dailey}, {Eisenhardt}, {Fabinsky},
  {Fajardo-Acosta}, {Fowler}, {Gelino}, {Grillmair}, {Heinrichsen}, {Kendall},
  {Kirkpatrick}, {Liu}, {Masci}, {McCallon}, {Nugent}, {Papin}, {Rice},
  {Royer}, {Ryan}, {Sevilla}, {Sonnett}, {Stevenson}, {Thompson}, {Wheelock},
  {Wiemer}, {Wittman}, {Wright}, \& {Yan}}]{mainzer2014}
{Mainzer}, A., {Bauer}, J., {cutri}, R.~M., {et~al.} 2014, \apj, 792, 30

\bibitem[{{Mainzer} {et~al.}(2011){Mainzer}, {Bauer}, {Grav}, {Masiero},
  {cutri}, {Dailey}, {Eisenhardt}, {McMillan}, {Wright}, {Walker}, {Jedicke},
  {Spahr}, {Tholen}, {Alles}, {Beck}, {Brandenburg}, {Conrow}, {Evans},
  {Fowler}, {Jarrett}, {Marsh}, {Masci}, {McCallon}, {Wheelock}, {Wittman},
  {Wyatt}, {DeBaun}, {Elliott}, {Elsbury}, {Gautier}, {Gomillion}, {Leisawitz},
  {Maleszewski}, {Micheli}, \& {Wilkins}}]{mainzer2011}
{Mainzer}, A., {Bauer}, J., {Grav}, T., {et~al.} 2011, \apj, 731, 53

\bibitem[{{Marton} {et~al.}(2019){Marton}, {{\'A}brah{\'a}m}, {Szegedi-Elek},
  {Varga}, {Kun}, {K{\'o}sp{\'a}l}, {Varga-Vereb{\'e}lyi}, {Hodgkin},
  {Szabados}, {Beck}, \& {Kiss}}]{Marton2019}
{Marton}, G., {{\'A}brah{\'a}m}, P., {Szegedi-Elek}, E., {et~al.} 2019, \mnras,
  487, 2522

\bibitem[{{Masci} {et~al.}(2019){Masci}, {Laher}, {Rusholme}, {Shupe}, {Groom},
  {Surace}, {Jackson}, {Monkewitz}, {Beck}, {Flynn}, {Terek}, {Landry},
  {Hacopians}, {Desai}, {Howell}, {Brooke}, {Imel}, {Wachter}, {Ye}, {Lin},
  {Cenko}, {Cunningham}, {Rebbapragada}, {Bue}, {Miller}, {Mahabal}, {Bellm},
  {Patterson}, {Juri{\'c}}, {Golkhou}, {Ofek}, {Walters}, {Graham}, {Kasliwal},
  {Dekany}, {Kupfer}, {Burdge}, {Cannella}, {Barlow}, {Van Sistine}, {Giomi},
  {Fremling}, {Blagorodnova}, {Levitan}, {Riddle}, {Smith}, {Helou}, {Prince},
  \& {Kulkarni}}]{masci2019}
{Masci}, F.~J., {Laher}, R.~R., {Rusholme}, B., {et~al.} 2019, \pasp, 131,
  018003

\bibitem[{{Megeath} {et~al.}(2012){Megeath}, {Gutermuth}, {Muzerolle},
  {Kryukova}, {Flaherty}, {Hora}, {Allen}, {Hartmann}, {Myers}, {Pipher},
  {Stauffer}, {Young}, \& {Fazio}}]{meg12}
{Megeath}, S.~T., {Gutermuth}, R., {Muzerolle}, J., {et~al.} 2012, \aj, 144,
  192

\bibitem[{{Merc} {et~al.}(2020){Merc}, {G{\'a}lis}, {K{\'a}ra}, {Wolf}, \&
  {Vra{\v{s}}{\v{t}}{\'a}k}}]{Merc2020}
{Merc}, J., {G{\'a}lis}, R., {K{\'a}ra}, J., {Wolf}, M., \&
  {Vra{\v{s}}{\v{t}}{\'a}k}, M. 2020, \mnras, 499, 2116

\bibitem[{{Meyer} {et~al.}(1997){Meyer}, {Calvet}, \& {Hillenbrand}}]{mey97}
{Meyer}, M.~R., {Calvet}, N., \& {Hillenbrand}, L.~A. 1997, \aj, 114, 288

\bibitem[{{Munari} {et~al.}(2022){Munari}, {Alcal{\'a}}, {Frasca}, {Masetti},
  {Traven}, {Akras}, \& {Zampieri}}]{Munari2022}
{Munari}, U., {Alcal{\'a}}, J.~M., {Frasca}, A., {et~al.} 2022, \aap, 661, A124

\bibitem[{{Mu{\v{z}}i{\'c}} {et~al.}(2022){Mu{\v{z}}i{\'c}}, {Almendros-Abad},
  {Bouy}, {Kubiak}, {Pe{\~n}a Ram{\'\i}rez}, {Krone-Martins}, {Moitinho}, \&
  {Concei{\c{c}}{\~a}o}}]{Muzic2022}
{Mu{\v{z}}i{\'c}}, K., {Almendros-Abad}, V., {Bouy}, H., {et~al.} 2022, \aap,
  668, A19

\bibitem[{{Nagy} {et~al.}(2023){Nagy}, {Park}, {{\'A}brah{\'a}m},
  {K{\'o}sp{\'a}l}, {Cruz-S{\'a}enz de Miera}, {Kun}, {Siwak}, {Szab{\'o}},
  {Szil{\'a}gyi}, {Fiorellino}, {Giannini}, {Lee}, {Lee}, {Marton}, {Szabados},
  {Vitali}, {Andrzejewski}, {Gromadzki}, {Hodgkin}, {Jab{\l}o{\'n}ska},
  {Mendez}, {Merc}, {Michniewicz}, {Miko{\l}ajczyk}, {Pylypenko}, {Ratajczak},
  {Wyrzykowski}, {Zejmo}, \& {Zieli{\'n}ski}}]{nagy2023}
{Nagy}, Z., {Park}, S., {{\'A}brah{\'a}m}, P., {et~al.} 2023, \mnras, 524, 3344

\bibitem[{{Park} {et~al.}(2022){Park}, {K{\'o}sp{\'a}l}, {{\'A}brah{\'a}m},
  {Cruz-S{\'a}enz de Miera}, {Fiorellino}, {Siwak}, {Nagy}, {Giannini},
  {Carini}, {Szab{\'o}}, {Lee}, {Lee}, {Vitali}, {Kun}, {Cseh}, {Krezinger},
  {Kriskovics}, {Ordasi}, {P{\'a}l}, {Szak{\'a}ts}, {Vida}, \&
  {Vink{\'o}}}]{Park2022}
{Park}, S., {K{\'o}sp{\'a}l}, {\'A}., {{\'A}brah{\'a}m}, P., {et~al.} 2022,
  \apj, 941, 165

\bibitem[{{Piascik} {et~al.}(2014){Piascik}, {Steele}, {Bates}, {Mottram},
  {Smith}, {Barnsley}, \& {Bolton}}]{Piascik2014}
{Piascik}, A.~S., {Steele}, I.~A., {Bates}, S.~D., {et~al.} 2014, in Society of
  Photo-Optical Instrumentation Engineers (SPIE) Conference Series, Vol. 9147,
  Ground-based and Airborne Instrumentation for Astronomy V, ed. S.~K.
  {Ramsay}, I.~S. {McLean}, \& H.~{Takami}, 91478H

\bibitem[{{Prisinzano} {et~al.}(2022){Prisinzano}, {Damiani}, {Sciortino},
  {Flaccomio}, {Guarcello}, {Micela}, {Tognelli}, {Jeffries}, \&
  {Alcal{\'a}}}]{Prisinzano2022}
{Prisinzano}, L., {Damiani}, F., {Sciortino}, S., {et~al.} 2022, \aap, 664,
  A175

\bibitem[{{Ramos Almeida} {et~al.}(2009){Ramos Almeida}, {P{\'e}rez
  Garc{\'\i}a}, \& {Acosta-Pulido}}]{Ramos2009}
{Ramos Almeida}, C., {P{\'e}rez Garc{\'\i}a}, A.~M., \& {Acosta-Pulido}, J.~A.
  2009, \apj, 694, 1379

\bibitem[{{Randich} {et~al.}(2018){Randich}, {Tognelli}, {Jackson}, {Jeffries},
  {Degl'Innocenti}, {Pancino}, {Re Fiorentin}, {Spagna}, {Sacco}, {Bragaglia},
  {Magrini}, {Prada Moroni}, {Alfaro}, {Franciosini}, {Morbidelli},
  {Roccatagliata}, {Bouy}, {Bravi}, {Jim{\'e}nez-Esteban}, {Jordi}, {Zari},
  {Tautvai{\v{s}}iene}, {Drazdauskas}, {Mikolaitis}, {Gilmore}, {Feltzing},
  {Vallenari}, {Bensby}, {Koposov}, {Korn}, {Lanzafame}, {Smiljanic}, {Bayo},
  {Carraro}, {Costado}, {Heiter}, {Hourihane}, {Jofr{\'e}}, {Lewis}, {Monaco},
  {Prisinzano}, {Sbordone}, {Sousa}, {Worley}, \& {Zaggia}}]{Randich2018}
{Randich}, S., {Tognelli}, E., {Jackson}, R., {et~al.} 2018, \aap, 612, A99

\bibitem[{{Shu}(1977)}]{shu77}
{Shu}, F.~H. 1977, \apj, 214, 488

\bibitem[{{Siess} {et~al.}(2000){Siess}, {Dufour}, \& {Forestini}}]{sie00}
{Siess}, L., {Dufour}, E., \& {Forestini}, M. 2000, \aap, 358, 593

\bibitem[{{Siwak} {et~al.}(2023){Siwak}, {Hillenbrand}, {K{\'o}sp{\'a}l},
  {{\'A}brah{\'a}m}, {Giannini}, {De}, {Mo{\'o}r}, {Szil{\'a}gyi},
  {Jan{\'\i}k}, {Koen}, {Park}, {Nagy}, {Cruz-S{\'a}enz de Miera},
  {Fiorellino}, {Marton}, {Kun}, {Lucas}, {Udalski}, \&
  {Szab{\'o}}}]{Siwak2023}
{Siwak}, M., {Hillenbrand}, L.~A., {K{\'o}sp{\'a}l}, {\'A}., {et~al.} 2023,
  \mnras, 524, 5548

\bibitem[{{Siwak} {et~al.}(2018){Siwak}, {Ogloza}, {Moffat}, {Matthews},
  {Rucinski}, {Kallinger}, {Kuschnig}, {Cameron}, {Weiss}, {Rowe}, {Guenther},
  \& {Sasselov}}]{siwak2018}
{Siwak}, M., {Ogloza}, W., {Moffat}, A. F.~J., {et~al.} 2018, \mnras, 478, 758

\bibitem[{{Szab{\'o}} {et~al.}(2021){Szab{\'o}}, {K{\'o}sp{\'a}l},
  {{\'A}brah{\'a}m}, {Park}, {Siwak}, {Green}, {Mo{\'o}r}, {P{\'a}l},
  {Acosta-Pulido}, {Lee}, {Cseh}, {Cs{\"o}rnyei}, {Hanyecz},
  {K{\"o}nyves-T{\'o}th}, {Krezinger}, {Kriskovics}, {Ordasi}, {S{\'a}rneczky},
  {Seli}, {Szak{\'a}ts}, {Szing}, \& {Vida}}]{Szabo2021}
{Szab{\'o}}, Z.~M., {K{\'o}sp{\'a}l}, {\'A}., {{\'A}brah{\'a}m}, P., {et~al.}
  2021, \apj, 917, 80

\bibitem[{{Szab{\'o}} {et~al.}(2022){Szab{\'o}}, {K{\'o}sp{\'a}l},
  {{\'A}brah{\'a}m}, {Park}, {Siwak}, {Green}, {P{\'a}l}, {Acosta-Pulido},
  {Lee}, {Ibrahimov}, {Grankin}, {Kov{\'a}cs}, {Bora}, {B{\'o}di}, {Cseh},
  {Cs{\"o}rnyei}, {Dr{\'o}{\.z}d{\.z}}, {Hanyecz}, {Ign{\'a}cz}, {Kalup},
  {K{\"o}nyves-T{\'o}th}, {Krezinger}, {Kriskovics}, {Og{\l}oza}, {Ordasi},
  {S{\'a}rneczky}, {Seli}, {Szak{\'a}ts}, {S{\'o}dor}, {Szing}, {Vida}, \&
  {Vink{\'o}}}]{Szabo2022}
{Szab{\'o}}, Z.~M., {K{\'o}sp{\'a}l}, {\'A}., {{\'A}brah{\'a}m}, P., {et~al.}
  2022, \apj, 936, 64

\bibitem[{{Szegedi-Elek} {et~al.}(2020){Szegedi-Elek}, {{\'A}brah{\'a}m},
  {Wyrzykowski}, {Kun}, {K{\'o}sp{\'a}l}, {Chen}, {Marton}, {Mo{\'o}r}, {Kiss},
  {P{\'a}l}, {Szabados}, {Varga}, {Varga-Vereb{\'e}lyi}, {Andreas}, {Bachelet},
  {Bischoff}, {B{\'o}di}, {Breedt}, {Burgaz}, {Butterley}, {Carrasco},
  {{\v{C}}epas}, {Damljanovic}, {Gezer}, {Godunova}, {Gromadzki}, {Gurgul},
  {Hardy}, {Hildebrandt}, {Hoffmann}, {Hundertmark}, {Ihanec}, {Janulis},
  {Kalup}, {Kaczmarek}, {K{\"o}nyves-T{\'o}th}, {Krezinger}, {Kruszy{\'n}ska},
  {Littlefair}, {Maskoli{\={u}}nas}, {M{\'e}sz{\'a}ros}, {Miko{\l}ajczyk},
  {Mugrauer}, {Netzel}, {Ordasi}, {Pak{\v{s}}tien{\.{e}}}, {Rybicki},
  {S{\'a}rneczky}, {Seli}, {Simon}, {{\v{S}}i{\v{s}}kauskait{\.{e}}},
  {S{\'o}dor}, {Sokolovsky}, {Stenglein}, {Street}, {Szak{\'a}ts}, {Tomasella},
  {Tsapras}, {Vida}, {Zdanavi{\v{c}}ius}, {Zieli{\'n}ski}, {Zieli{\'n}ski}, \&
  {Zi{\'o}{\l}kowska}}]{Szegedi-Elek2020}
{Szegedi-Elek}, E., {{\'A}brah{\'a}m}, P., {Wyrzykowski}, {\L}., {et~al.} 2020,
  \apj, 899, 130

\bibitem[{{Tody}(1993)}]{Tody1993}
{Tody}, D. 1993, Astronomical Society of the Pacific Conference Series,
  Vol.~52, {IRAF in the Nineties}, ed. R.~J. {Hanisch}, R.~J.~V. {Brissenden},
  \& J.~{Barnes}, 173

\bibitem[{{Tognelli} {et~al.}(2018){Tognelli}, {Prada Moroni}, \&
  {Degl'Innocenti}}]{Tognelli2018}
{Tognelli}, E., {Prada Moroni}, P.~G., \& {Degl'Innocenti}, S. 2018, \mnras,
  476, 27

\bibitem[{{Tognelli} {et~al.}(2020){Tognelli}, {Prada Moroni},
  {Degl'Innocenti}, {Salaris}, \& {Cassisi}}]{Tognelli2020}
{Tognelli}, E., {Prada Moroni}, P.~G., {Degl'Innocenti}, S., {Salaris}, M., \&
  {Cassisi}, S. 2020, \aap, 638, A81

\bibitem[{{Vacca} {et~al.}(2003){Vacca}, {Cushing}, \& {Rayner}}]{Vacca2003}
{Vacca}, W.~D., {Cushing}, M.~C., \& {Rayner}, J.~T. 2003, \pasp, 115, 389

\bibitem[{{Wright} {et~al.}(2010){Wright}, {Eisenhardt}, {Mainzer}, {Ressler},
  {cutri}, {Jarrett}, {Kirkpatrick}, {Padgett}, {McMillan}, {Skrutskie},
  {Stanford}, {Cohen}, {Walker}, {Mather}, {Leisawitz}, {Gautier}, {McLean},
  {Benford}, {Lonsdale}, {Blain}, {Mendez}, {Irace}, {Duval}, {Liu}, {Royer},
  {Heinrichsen}, {Howard}, {Shannon}, {Kendall}, {Walsh}, {Larsen}, {Cardon},
  {Schick}, {Schwalm}, {Abid}, {Fabinsky}, {Naes}, \& {Tsai}}]{wright2010}
{Wright}, E.~L., {Eisenhardt}, P. R.~M., {Mainzer}, A.~K., {et~al.} 2010, \aj,
  140, 1868

\bibitem[{{Zsidi} {et~al.}(2022){Zsidi}, {Fiorellino}, {K{\'o}sp{\'a}l},
  {{\'A}brah{\'a}m}, {B{\'o}di}, {Hussain}, {Manara}, \&
  {P{\'a}l}}]{zsidi2022vw}
{Zsidi}, G., {Fiorellino}, E., {K{\'o}sp{\'a}l}, {\'A}., {et~al.} 2022, \apj,
  941, 177

\end{thebibliography}

\appendix


\section{Photometry}
\label{app:photometry}

Tables \ref{tab:photometry} and \ref{tab:photometryNIR} collect the optical and NIR photometry of Gaia18cjb, respectively.

\begin{table*}
 \center
 \caption{\label{tab:photometry}The optical photometry of Gaia18cjb we acquired and/or reduced. EFOSC $r$ photometry was taken with Bessell $R$ filter.}
\resizebox{\textwidth}{!}{%
  \begin{tabular}{lllccccc}
  \hline
  \hline
Instr. & Date & HJD & $B$ & $g$  & $V$  & $r$ & $i$ \\
     &    & & mag & mag & mag & mag & mag    \\
    \hline
PanSTaRRS & 2011 Oct 25 & 2455860.13  &  $-$  &  $-$             & $-$ &  $ 20.2 \pm 0.1$  & $-$ \\
PanSTaRRS & 2011 Oct 25 & 2455860.14  &  $-$  &  $-$             & $-$ &  $ 20.1 \pm 0.1$  & $-$ \\
PanSTaRRS & 2011 Nov 01 & 2455566.95  &  $-$  &  $22.1 \pm 0.2$  & $-$ &  $-$              & $-$ \\			          
PanSTaRRS & 2011 Nov 01 & 2455566.96  &  $-$  &  $22.0 \pm 0.2$  & $-$ &  $-$              & $-$ \\                         
PanSTaRRS & 2011 Nov 11 & 2455876.06  &  $-$  &  $-$             & $-$ &  $-$              & $19.2 \pm 0.1$ \\
PanSTaRRS & 2011 Nov 11 & 2455876.07  &  $-$  &  $-$             & $-$ &  $-$              & $19.4 \pm 0.1$ \\
PanSTaRRS & 2012 Jan 04 & 2455930.94  &  $-$  &  $-$             & $-$ &  $-$              & $19.1 \pm 0.1$ \\
PanSTaRRS & 2012 Jan 04 & 2455930.96  &  $-$  &  $-$             & $-$ &  $-$              & $19.3 \pm 0.1$ \\
PanSTaRRS & 2012 Oct 14 & 2456215.08  &  $-$  &  $21.8 \pm 0.2$  & $-$ &  $-$              & $-$ \\                           
PanSTaRRS & 2012 Oct 14 & 2456215.08  &  $-$  &  $21.7 \pm 0.2$  & $-$ &  $-$              & $-$ \\                           
PanSTaRRS & 2012 Oct 14 & 2456215.10  &  $-$  &  $22.4 \pm 0.2$  & $-$ &  $-$              & $-$ \\                           
PanSTaRRS & 2012 Oct 14 & 2456215.10  &  $-$  &  $21.7 \pm 0.2$  & $-$ &  $-$              & $-$ \\                           
PanSTaRRS & 2012 Oct 19 & 2456220.09  &  $-$  &  $-$             & $-$ &  $ 20.1 \pm 0.1$  & $-$ \\
PanSTaRRS & 2012 Oct 19 & 2456220.10  &  $-$  &  $-$             & $-$ &  $ 20.0 \pm 0.1$  & $18.9 \pm 0.1$ \\
PanSTaRRS & 2012 Oct 19 & 2456220.11  &  $-$  &  $-$             & $-$ &  $ 20.3 \pm 0.1$  & $19.0 \pm 0.1$ \\
PanSTaRRS & 2012 Nov 08 & 2456240.09  &  $-$  &  $-$             & $-$ &  $ 19.7 \pm 0.1$ & $-$ \\
PanSTaRRS & 2012 Nov 08 & 2456240.09  &  $-$  &  $-$             & $-$ &  $ 20.1 \pm 0.1$ & $-$ \\
PanSTaRRS & 2012 Nov 08 & 2456240.10  &  $-$  &  $-$             & $-$ &  $ 19.8 \pm 0.1$ & $-$ \\
PanSTaRRS & 2012 Nov 08 & 2456240.10  &  $-$  &  $-$             & $-$ &  $ 20.1 \pm 0.1$ & $-$ \\
PanSTaRRS & 2012 Nov 09 & 2456241.05  &  $-$  &  $-$             & $-$ &  $ 20.1 \pm 0.1$ & $-$ \\
PanSTaRRS & 2012 Nov 09 & 2456241.06  &  $-$  &  $-$             & $-$ &  $ 20.0 \pm 0.1$ & $-$ \\
PanSTaRRS & 2012 Nov 11 & 2456243.04  &  $-$  &  $21.6 \pm 0.2$  & $-$ &  $-$              & $-$ \\                           
PanSTaRRS & 2012 Nov 11 & 2456243.05  &  $-$  &  $21.1 \pm 0.2$  & $-$ &  $-$              & $-$ \\                           
PanSTaRRS & 2012 Dec 23 & 2456284.96  &  $-$  &  $-$             & $-$ &  $-$              & $19.0 \pm 0.1$ \\
PanSTaRRS & 2013 Mar 05 & 2456356.78  &  $-$  &  $21.9 \pm 0.2$  & $-$ &  $-$              & $-$ \\                           
PanSTaRRS & 2013 Mar 05 & 2456356.78  &  $-$  &  $22.1 \pm 0.2$  & $-$ &  $-$              & $-$ \\                           
PanSTaRRS & 2013 Mar 05 & 2456356.79  &  $-$  &  $22.6 \pm 0.2$  & $-$ &  $-$              & $-$ \\                           
PanSTaRRS & 2013 Mar 05 & 2456356.90  &  $-$  &  $21.8 \pm 0.2$  & $-$ &  $-$              & $-$ \\                           
PanSTaRRS & 2013 Dec 04 & 2456631.01  &  $-$  &  $21.4 \pm 0.2$  & $-$ &  $ 19.9 \pm 0.1$ & $-$ \\                
PanSTaRRS & 2013 Dec 04 & 2456631.02  &  $-$  &  $21.5 \pm 0.2$  & $-$ &  $ 19.8 \pm 0.1$ & $-$ \\              
PanSTaRRS & 2013 Dec 23 & 2456649.99  &  $-$  &  $-$             & $-$ &  $-$             & $19.0 \pm 0.1$ \\
PanSTaRRS & 2013 Dec 24 & 2456650.00  &  $-$  &  $-$             & $-$ &  $-$             & $19.1 \pm 0.1$ \\
Adiyaman & 2020 Nov 13 & 2459167.49 & -- & 17.688$\pm$0.052 & -- & 16.227$\pm$0.026 & 15.564$\pm$0.060 \\
Adiyaman & 2020 Nov 27 & 2459180.51 & -- & 17.617$\pm$0.065 & -- & 16.149$\pm$0.025 & 15.495$\pm$0.054 \\
Adiyaman & 2020 Nov 29 & 2459183.39 & -- & 17.425$\pm$0.082 & -- & 16.162$\pm$0.031 & 15.491$\pm$0.045 \\
Adiyaman & 2021 Jan 10 & 2459225.28 & -- & 17.498$\pm$0.073 & -- & 16.141$\pm$0.036 & 15.504$\pm$0.047 \\
Adiyaman & 2021 Jan 24 & 2459239.45 & -- & 17.575$\pm$0.086 & -- & 16.213$\pm$0.033 & 15.503$\pm$0.052 \\
Adiyaman & 2021 Feb 10 & 2459256.40 & -- & 17.613$\pm$0.058 & -- & 16.174$\pm$0.025 & 15.503$\pm$0.056 \\
EFOSC & 2021 May 09 & 2459344.49 & 18.220$\pm$0.022 & -- & 16.778$\pm$0.030 & 16.028$\pm$0.021 & --  \\ 
RC80 & 2020 Oct 22 & 2459145.56 & 18.32$\pm$0.18 & 17.597$\pm$0.035 & 16.903$\pm$0.032 & 16.242$\pm$0.014 & 15.516$\pm$0.012 \\
RC80 & 2020 Nov 05 & 2459159.67 & 18.21$\pm$0.12 & 17.390$\pm$0.076 & 16.832$\pm$0.036 & 16.180$\pm$0.021 & 15.450$\pm$0.025 \\
RC80 & 2020 Nov 22 & 2459175.65 & 18.50$\pm$0.15 & 17.562$\pm$0.038 & 16.880$\pm$0.041 & 16.213$\pm$0.012 & 15.470$\pm$0.011 \\
RC80 & 2021 Feb 15 & 2459261.26 & 18.27$\pm$0.15 & 17.520$\pm$0.044 & 16.802$\pm$0.055 & 16.162$\pm$0.029 & 15.463$\pm$0.027 \\
RC80 & 2021 Feb 18 & 2459264.27 & 18.17$\pm$0.11 & 17.511$\pm$0.079 & 16.830$\pm$0.051 & 16.167$\pm$0.031 & 15.461$\pm$0.030 \\
RC80 & 2021 Feb 20 & 2459266.25 & 18.49$\pm$0.14 & 17.605$\pm$0.048 & 16.896$\pm$0.051 & 16.217$\pm$0.035 & 15.503$\pm$0.030 \\
RC80 & 2021 Feb 23 & 2459269.25 & 17.80$\pm$0.25 & 17.29$\pm$0.15 & 16.54$\pm$0.14 & 16.20$\pm$0.13 & 15.296$\pm$0.088 \\
RC80 & 2021 Feb 24 & 2459270.27 & 18.57$\pm$0.24 & 17.570$\pm$0.083 & 16.846$\pm$0.083 & 16.11$\pm$0.15 & 15.436$\pm$0.047\\
RC80 & 2021 Sep 08 & 2459465.61 & 18.30$\pm$0.13 & 17.530$\pm$0.053 & 16.785$\pm$0.032 & 16.137$\pm$0.016 & 15.410$\pm$0.018 \\
RC80 & 2021 Oct 20 & 2459507.61 & 18.181$\pm$0.096 & -- & 16.86$\pm$0.16 & -- & -- \\
RC80 & 2021 Oct 23 & 2459510.52 & -- & 17.548$\pm$0.052 & -- & 16.128$\pm$0.021 & 15.415$\pm$0.016 \\
RC80 & 2021 Nov 24 & 2459542.61 & 18.204$\pm$0.080 & 17.514$\pm$0.037 & 16.781$\pm$0.037 & 16.118$\pm$0.015 & 15.414$\pm$0.016 \\
RC80 & 2022 Jan 07 & 2459586.59 & 18.19$\pm$0.10 & 17.464$\pm$0.047 & 16.688$\pm$0.036 & 16.013$\pm$0.017 & 15.319$\pm$0.017 \\
RC80 & 2022 Feb 05 & 2459616.25 & 18.168$\pm$0.075 & 17.331$\pm$0.030 & 16.637$\pm$0.029 & 15.983$\pm$0.016 & 15.265$\pm$0.011 \\
RC80 & 2022 Sep 24 & 2459846.63 & 18.16$\pm$0.11 & 17.313$\pm$0.034 & 16.636$\pm$0.030 & 15.982$\pm$0.026 & 15.290$\pm$0.016 \\
RC80 & 2022 Oct 25 & 2459878.49 & 17.910$\pm$0.091 & 17.304$\pm$0.034 & 16.588$\pm$0.030 & 15.944$\pm$0.015 & 15.250$\pm$0.012 \\
RC80 & 2022 Nov 25 & 2459908.62 & 18.172$\pm$0.085 & 17.478$\pm$0.032 & 16.731$\pm$0.029 & 16.090$\pm$0.015 & 15.405$\pm$0.016 \\
RC80 & 2022 Dec 24 & 2459938.47 & 18.09$\pm$0.13 & 17.316$\pm$0.039 & 16.649$\pm$0.038 & 15.998$\pm$0.019 & 15.290$\pm$0.013 \\
RC80 & 2023 Feb 06 & 2459982.34 & 18.06$\pm$0.19 & 17.363$\pm$0.094 & 16.599$\pm$0.056 & 15.933$\pm$0.023 & 15.220$\pm$0.017 \\
RC80 & 2023 Mar 22 & 2460026.36 & 18.10$\pm$0.16 & 17.261$\pm$0.058 & 16.517$\pm$0.044 & 15.911$\pm$0.026 & 15.210$\pm$0.015 \\
Suhora & 2021 Dec 13 & 2459561.56 & 18.267$\pm$0.096 & 17.340$\pm$0.026 & 16.576$\pm$0.025 & 15.924$\pm$0.022 & 15.235$\pm$0.082 \\ 
Suhora & 2021 Jan 03 & 2459218.46 & -- & 17.499$\pm$0.045 & -- & 16.092$\pm$0.024 & 15.422$\pm$0.067\\ 
Suhora & 2021 Jan 22 & 2459237.46 & -- & -- & 16.782$\pm$0.042 & -- & -- \\ 
\hline
\hline
  \end{tabular}
  }
\end{table*}


\begin{table*}
 \center
 \caption{\label{tab:photometryNIR}The near-infrared photometry of Gaia18cjb we reduced.}
  \begin{tabular}{lllccccc}
  \hline
  \hline
Instr. & Date & HJD & $Y$ & $z$ & $J$ & $H$ & $K$  \\
     &    & & mag & mag & mag & mag & mag\\
    \hline
UKIDSS    & 2007 Apr 15 & 2454205.5 & $ - $          & $-$            & $-$              & $-$              & 12.620$\pm$0.040 \\ 
UKIDSS    & 2010 Oct 24 & 2455493.5 & $ - $          & $-$            & 16.210$\pm$0.040 & 14.860$\pm$0.050 & 13.260$\pm$0.040  \\
PanSTaRRS & 2010 Feb 15 & 2455242.8 & $18.7 \pm 0.1$ & $-$            & $-$              & $-$              & $-$ \\
PanSTaRRS & 2010 Feb 15 & 2455242.8 & $18.3 \pm 0.1$ & $-$            & $-$              & $-$              & $-$ \\
PanSTaRRS & 2010 Feb 15 & 2455242.8 & $18.6 \pm 0.1$ & $-$            & $-$              & $-$              & $-$ \\
PanSTaRRS & 2010 Feb 15 & 2455242.8 & $18.1 \pm 0.1$ & $-$            & $-$              & $-$              & $-$ \\
PanSTaRRS & 2010 Oct 22 & 2455492.1 & $18.2 \pm 0.1$ & $-$            & $-$              & $-$              & $-$ \\
PanSTaRRS & 2010 Oct 22 & 2455492.1 & $18.1 \pm 0.1$ & $-$            & $-$              & $-$              & $-$ \\
PanSTaRRS & 2011 Sep 18 & 2455823.1 & $18.2 \pm 0.1$ & $-$            & $-$              & $-$              & $-$ \\
PanSTaRRS & 2011 Sep 18 & 2455823.1 & $17.9 \pm 0.1$ & $-$            & $-$              & $-$              & $-$ \\
PanSTaRRS & 2011 Oct 22 & 2455856.8 & $18.1 \pm 0.1$ & $18.6 \pm 0.1$ & $-$              & $-$              & $-$ \\
PanSTaRRS & 2011 Oct 22 & 2455256.8 & $-$            & $18.4 \pm 0.1$ & $-$              & $-$              & $-$ \\
PanSTaRRS & 2011 Oct 27 & 2455862.1 & $-$            & $18.5 \pm 0.1$ & $-$              & $-$              & $-$ \\
PanSTaRRS & 2011 Oct 27 & 2455862.1 & $-$            & $18.6 \pm 0.1$ & $-$              & $-$              & $-$ \\
PanSTaRRS & 2012 Feb 27 & 2455984.7 & $18.0 \pm 0.1$ & $-$            & $-$              & $-$              & $-$ \\  
PanSTaRRS & 2012 Feb 27 & 2455984.7 & $18.1 \pm 0.1$ & $-$            & $-$              & $-$              & $-$ \\
PanSTaRRS & 2012 Feb 28 & 2455485.1 & $-$            & $18.4 \pm 0.1$ & $-$              & $-$              & $-$ \\
PanSTaRRS & 2012 Feb 28 & 2455485.1 & $-$            & $18.5 \pm 0.1$ & $-$              & $-$              & $-$ \\
PanSTaRRS & 2012 Mar 20 & 2456006.8 & $-$            & $18.5 \pm 0.1$ & $-$              & $-$              & $-$ \\
PanSTaRRS & 2012 Mar 20 & 2456006.8 & $-$            & $18.6 \pm 0.1$ & $-$              & $-$              & $-$ \\
PanSTaRRS & 2012 Oct 22 & 2456223.1 & $-$            & $18.3 \pm 0.1$ & $-$              & $-$              & $-$ \\
PanSTaRRS & 2013 Oct 20 & 2456586.1 & $18.0 \pm 0.1$ & $18.2 \pm 0.1$ & $-$              & $-$              & $-$ \\
PanSTaRRS & 2014 Oct 08 & 2456939.1 & $18.0 \pm 0.1$ & $-$            & $-$              & $-$              & $-$ \\                                                        
SOFI      & 2021 May 08 & 2459343.0 & $-$            & $-$            & 13.505$\pm$0.058 & 12.720$\pm$0.035 & 11.911$\pm$0.042 \\
GTC       & 2022 Oct 10 & 2459862.5 & $-$            & $-$            & 13.354 $\pm$0.080 & 12.672$\pm$0.073 & 11.952$\pm$0.106  \\
GTC       & 2022 Oct 26 & 2459878.6 & $-$            & $-$            & 13.396 $\pm$0.124 & 12.675$\pm$0.067 & 12.059$\pm$0.160  \\
                                                                        
\hline
\hline
  \end{tabular}
\end{table*}

\begin{table}
 \center
 \caption{\label{tab:photometrySED}The photometry used for the SED of Gaia18cjb.}
  \begin{tabular}{llcc}
  \hline
  \hline
Instr. & Year & $\lambda$ & Flux \\
       &      & $\mu$m & mJy   \\
    \hline
WISE      & 2013 & 11.55 &  $ 93.1    \pm 1.3     $ \\
WISE      & 2013 & 22.09 &  $ 145.0   \pm 3.0     $ \\
PanSTaRRS & 2012 & 0.747 &  $ 0.06580 \pm 0.0014  $ \\
PanSTaRRS & 2012 & 0.865 &  $ 0.1300  \pm 0.0020  $ \\
PanSTaRRS & 2012 & 0.959 &  $ 0.1620  \pm 0.0040  $ \\
WISE      & 2011 & 3.350 &  $ 14.40   \pm 0.30    $ \\
WISE      & 2011 & 4.600 &  $ 37.50   \pm 0.70    $ \\
AKARI     & 2010 & 8.610 &  $ 114     \pm 16      $ \\
AKARI     & 2010 & 64.99 &  $ 461 	 \pm 76      $ \\
AKARI     & 2010 & 90.00 &  $ 506 	 \pm 81      $ \\
2MASS     & 2000 & 1.239 &  $ 0.665   \pm 0.052   $ \\
2MASS     & 2000 & 1.649 &  $ 1.430   \pm 0.080   $ \\
2MASS     & 2000 & 2.163 &  $ 3.95    \pm 0.15    $ \\
SDSS      & 1998 & 0.477 &  $ 0.00847 \pm 0.00037 $ \\
SDSS      & 1998 & 0.481 &  $ 0.00849 \pm 0.00037 $ \\
SDSS      & 1998 & 0.612 &  $ 0.03260 \pm 0.00040 $ \\
\hline
EFOSC & 2021 &  0.44   &  $1834 \pm 38$ \\ 
Suhora & 2021 &  0.52  &  $0.373 \pm 0.016$ \\ 
EFOSC & 2021 &  0.5476 &  $6331 \pm 177$ \\
EFOSC & 2021 &  0.6431 &  $14823 \pm 285$ \\
Suhora & 2021 &  0.67  &  $1.638 \pm 0.036$ \\
Suhora & 2021 &  0.79  &  $3.226 \pm 0.198$ \\
SOFI & 2021 & 1.247    &  $57.8 \pm 3.1$ \\ 
SOFI & 2021 & 1.653    &  $75.7 \pm 2.4$ \\
SOFI & 2021 & 2.162    &  $103.2 \pm 4.0$ \\
\hline
\hline
  \end{tabular}
\end{table}

\section{Spectra}
\label{app:spectra}

\begin{figure*}
    \includegraphics[width=\textwidth]{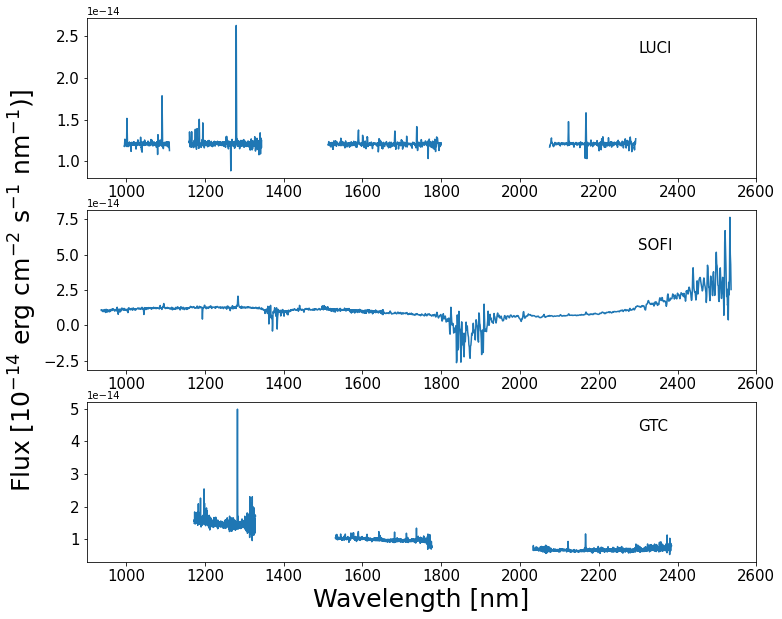}
    \label{fig:calibrated_spectra}
    \caption{Flux calibrated NIR spectra of Gaia18cjb.}
\end{figure*}

Figure\,\ref{fig:calibrated_spectra} shows the NIR flux calibrated spectra of Gaia18cjb.

\end{document}